\documentclass[aps,reprint,twocolumn,showpacs,floatfix,superscriptaddress]{revtex4-2}
\usepackage{amsmath,amssymb,color,braket,hyphenat,makeidx,xcolor}
\usepackage[ocgcolorlinks,colorlinks=true,linkcolor=blue,citecolor=red,linktocpage=true]{hyperref}

\usepackage{graphicx}
\usepackage[caption=false]{subfig}

\begin{document}
	%%%%%%%%%%%%%%%%%%%%%%%%%%%%%%%%%%%%%%%%%%%%%%%%%%%%%%%%%%%%%%%%%%%%%%%%%%%%%%%%%%%%%%%%%%%%%%%%%%%%
	\title{Coherence-Driven Quantum Battery Charging via Autonomous Thermal Machines: Energy Transfer, Memory Effects, and Ergotropy Enhancement}
	%%%%%%%%%%%%%%%%%%%%%%%%%%%%%%%%%%%%%%%%%%%%%%%%%%%%%%%%%%%%%%%%%%%%%%%%%%%%%%%%%%%%%%%%%%%%%%%%%%%%
	\author{A. Khoudiri}
	\email{khoudiri.achraf@etu.uae.ac.ma}
	\affiliation{Laboratory of R$\&$D in Engineering Sciences, Faculty of Sciences and Techniques Al-Hoceima, Abdelmalek Essaadi University, Tetouan,	Morocco.}
	%%%%%%%%%%%%%%%%%%%%%%%%%%%%%%%%%%%%%%%%%%%%%%%%%%%%%%%%%%%%%%%%%%%%%%%%%%%%%%%%%%%%%%%%%%%%%%%%%%%%
	\author{A. Oularabi}
	\email{oularabi.abderrahman@etu.uae.ac.ma}
	\affiliation{Laboratory of R$\&$D in Engineering Sciences, Faculty of Sciences and Techniques  Al-Hoceima, Abdelmalek Essaadi University, Tetouan,	Morocco.}
	%%%%%%%%%%%%%%%%%%%%%%%%%%%%%%%%%%%%%%%%%%%%%%%%%%%%%%%%%%%%%%%%%%%%%%%%%%%%%%%%%%%%%%%%%%%%%%%%%%%%
	\author{K. El Anouz}
	\email{kelanouz@uae.ac.ma}
	\affiliation{Laboratory of R$\&$D in Engineering Sciences, Faculty of Sciences and Techniques  Al-Hoceima, Abdelmalek Essaadi University, Tetouan,	Morocco.}
	%%%%%%%%%%%%%%%%%%%%%%%%%%%%%%%%%%%%%%%%%%%%%%%%%%%%%%%%%%%%%%%%%%%%%%%%%%%%%%%%%%%%%%%%%%%%%%%%%%%%
	\author{\.{I}. Demir}
	\email{idemir@cumhuriyet.edu.tr}
	\affiliation{Department of Nanotechnology Engineering, Sivas Cumhuriyet University, 58140 Sivas, T\"{u}rkiye}
	\affiliation{Sivas Cumhuriyet University Nanophotonics Application and Research Center-CUNAM, 58140 Sivas, T\"{u}rkiye
	}    
	%%%%%%%%%%%%%%%%%%%%%%%%%%%%%%%%%%%%%%%%%%%%%%%%%%%%%%%%%%%%%%%%%%%%%%%%%%%%%%%%%%%%%%%%%%%%%%%%%%%%
	\author{A. El Allati}
	\email{eabderrahim@uae.ac.ma}
	\affiliation{Laboratory of R$\&$D in Engineering Sciences, Faculty of Sciences and Techniques Al-Hoceima, Abdelmalek Essaadi University, Tetouan,	Morocco.}
	%%%%%%%%%%%%%%%%%%%%%%%%%%%%%%%%%%%%%%%%%%%%%%%%%%%%%%%%%%%%%%%%%%%%%%%%%%%%%%%%%%%%%%%%%%%%%%%%%%%%    
	\begin{abstract}
		In this work, we study a hybrid quantum system composed of a quantum battery and a coherence-driven charger interacting with a Quantum Autonomous Thermal Machine (QATM). The QATM, made of two qubits, each coupled to Markovian bosonic thermal reservoirs at different temperatures, acts as a structured environment that mediates energy and coherence between the charger and the battery. By applying a coherent driving field on the charger, we investigate the coherence injection effect on the dynamics, including non-Markovianity, power of charging, coherence storage, and ergotropy. We show that the QATM filters the decoherence induced by the thermal baths and induces non-Markovian memory effects due to correlation backflow. Our results demonstrate that coherence driving enhances the battery's ergotropy; coherence driving raises the maximum ergotropy by approximately 40\% compared to the case without coherence driving, and the power of charging by preserving the internal energy of the charger.

	\end{abstract}
	\maketitle
	%---------------------------------------------------------------------------------
	\section{Introduction}
	%---------------------------------------------------------------------------------
	There is a worldwide interest in using quantum phenomena to develop the quantum technology framework \cite{intro1,intro2}, especially in developing batteries at the quantum level \cite{intro3, intro4, intro5}. Moreover, it has been shown that the relationship between quantum batteries and quantum-scale energy storage plays an important role in quantum thermodynamics. Using autonomous thermal machines (QATMs) in quantum thermodynamics by taking advantage of quantum connections, non-Markovianity, the size of Hilbert space, and quantum coherence as resources has helped improve and speed up the charging process of quantum batteries \cite{intro6,intro7, intro8, intro9, intro10,intro10a,into10b,into10c,intro10aa,intro10aaa}.\\
	
	Numerous works have been introduced to study quantum batteries and the charging process based on energy transfer between qubits or the coherence-driven charging process under the effect of non-Markovian channels. For example, we can cite the role of coherence and correlation in the charging process \cite{intro11, intro11a,intro12,intro13}. In the context of autonomous machines, QATMs present many useful tools for exploiting reservoir temperature as a working generator. They can be used as a resource to filter the interaction between external systems and their environments as well as their role in generating entanglement, coherence and work \cite{intro14, intro15, intro16, intro17}. Specifically, if the memory effects appear in the dynamics of a thermal machine cycle, then the maximum power and efficiency of the thermal machine over time are extremely significant \cite{intro18, intro19}. In this inspiration, the following questions arise: What is the QATM's effect on the charging process of the quantum battery? What is the role of Hilbert space dimension, correlation, and non-Markovianity in the dynamics, control of charging, and transfer of energy and coherence in this context?\\
	
	Coherence-driven charging is a mechanism that injects coherence using an external field \cite{intro19a,intro20,intro21,intro22}. It accelerates and strengthens the charging process, besides, it also increases the efficiency of work extraction compared to conventional (incoherent) charging mechanisms. However, the energy transfer mediated by qubits plays an important role in transferring energy between qubits and the charging process over time. The charging is related to energy transfer using an ancillary system or the effect of electrodynamics between different systems, qubits or harmonic oscillators \cite{intro23, intro24}. Moreover, the impact of correlation on the charging process, energy transfer, and extracted work over time in a two-level system quantum charger-battery is well discussed in \cite{intro25}. The energy transfer based on an auxiliary system and coherence-driven charging is investigated in Ref.\cite{intro26}. In this regard, our main objective is to develop a study highlighting the impact of QATM on the coherence-driven charging process \cite{intro22,intro24,intro25,intro26}. Then, the capacity of QATM to charge the quantum battery in the presence of memory effects imposed by the autonomous quantum thermal machine will also be examined. This aligns with No-Go theorem in the context of closed quantum systems \cite{intro26a}.\\
	
	The impact of an external field is evaluated on energy transfer, storage, ergotropy, and coherence transfer in the presence of the memory effect imposed by a QATM. Moreover, the effect of QATM in the presence and absence of coherence driving for the quantum charger is examined. The model is structured as follows: two Markovian cold and hot reservoirs, namely \(R_1\) and \(R_2\), respectively, are coupled to two-qubit QATM \(M_{12}\) at thermal equilibrium. The composite system \(M_{12} = M_1 \otimes M_2\) represents the QATM, which is coupled to a quantum charger \(C\). Then, the external laser excitation induces coherence driving, which improves the charging process of the quantum battery, denoted as \(B\). Particularly, the impact of \(M_{12}\) on the evolution of charger and battery systems is shown. It is crucial to use the derivative of the trace distance to measure the memory effect between \(C\) and \(B\). The relative entropy of coherence is used to quantify the coherence of the charger and battery, respectively. Finally, the ability of quantum batteries to extract maximum work using quantum ergotropy in the presence and absence of the external field is also investigated.\\

	The rest of the paper is organized as follows: In Sec. \ref{sec:Model}, the theoretical model of the composite QATM, charger-field, and battery systems is described. In Sec. \ref{sec:Impact_of_QATM_Coherence-driving_on_chargig_process}, the impact of thermal machine and coherence-driving on the charging process is discussed. In Sec. \ref{sec:Impact_the_nonMarkovianity_on_chargig_process}, the impact of non-Markovianity on the power of charging and coherence transfer on the quantum battery over time is evaluated. Finally, all the obtained physical results are summarized in Sec. \ref{conc}.
	
	%---------------------------------------------------------------------------------
	\section{The dynamics of the composite QATM-Charger-Battery system}\label{sec:Model}
\begin{figure}[h!]
		%\centering
		\includegraphics[width=\columnwidth]{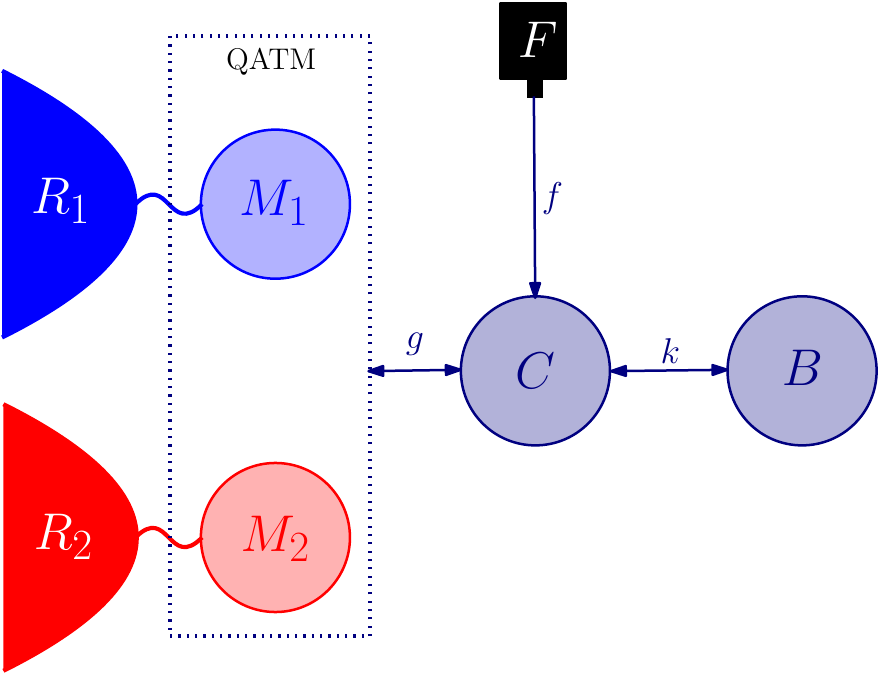}
		\makeatletter\long\def\@ifdim#1#2#3{#2}\makeatother
		\caption{Diagram of a QATM, namely $M_{12}=M_1\otimes M_2$ coupled to a quantum charger (C). The coherence driving is represented by a laser (F) which is coupled directly to (C), where $f$ is the field amplitude. Besides, (B) is the quantum battery coupled to the charger $C$, where $R_1$ and $R_2$ are the cold and hot bosonic reservoirs, respectively. Note that $g$ and $k$ are the coupling strengths between QATM-charger and battery-charger, respectively.}
		\label{f1}
	\end{figure}

	The proposed model is composed of three parts: two bosonic-Markovian reservoirs $R_1$ and $R_2$ of the inverse of temperatures $\beta_1$ and $\beta_2$, respectively, where $\beta_1\geq\beta_2$. The second part represents a QATM, $M_{12}$, which consists of two qubits $M_1$ and $M_2$ in their thermal equilibrium states. Each qubit of the QATM is coupled to its reservoir. The third part is composed of a charger (qubit) noted by $(C)$, which is excited using a quantum field employing a laser $(F)$. The quantum charger is coupled to a quantum battery (qubit) $(B)$, as shown in Fig.(\ref{f1}). 
	
	%%%%%%%%%%%%%%%%%%%%%%%%%%%%%%% Fig 1 %%%%%%%%%%%%%%%%%%%%%%%%%%%%%%%%%%%%%%%%%%%%%%%%%%
	%%%%%%%%%%%%%%%%%%%%%%%%%%%%%%%%%%%%%%%%%%%%%%%%%%%%%%%%%%%%%%%%%%%%%%%%%%%%%%%%%%%%%%%%%%%%%%%%%%%%%%%%%%%%%%%%%%%%%%
	%---------------------------------------------------------------------------------
	\subsection{Hamiltonian and dynamics}\label{sec:Hamiltonian_Model}
	The total Hamiltonian, already described above, is written in the following compact form 
	\begin{equation}
		\hat{H}=\sum_{m=1}^{2}\hat{H}_{R_{m}}+\hat{H}_{M_{12}}+\hat{H}_{R-M_{12}}+\hat{H}_{C-B}(t)+\hat{H}_{M_{12}-C},
	\end{equation}
	$\hbar=1$,  where
	\begin{equation}
		\hat{H}_{R_{m}}=\sum_k E_{mk}\hat{a}_{mk}^{\dagger}\hat{a}_{mk}, 
	\end{equation}
	is the free Hamiltonian of the bosonic reservoir $R_m$, which is modeled by a large set of harmonic oscillators, with $E_{mk}$ being the spacing energy.  $\hat{a}_{mk}$ $(\hat{a}^{\dagger}_{mk})$ are the annihilation and creation  operators of the oscillator $k$ for $m=\{1,2\}$. The Hamiltonian $\hat{H}_{M_{12}}$ of the QATM is given as 
	\begin{align}
		\hat{H}_{M_{12}}&=\sum_{m=1}^{2}\hat{H}_{M_{m}}\quad ,\quad
		\hat{H}_{M_{m}}&=\omega_{M_m}\hat{\sigma}_{M_m}^{+}\hat{\sigma}_{M_m}^{-},
	\end{align}
	$\hat{H}_{M_m}$ is the free Hamiltonian for the thermal machine qubit $M_m$ and $\omega_{M_m}$ denotes the spacing energy. $\hat{\sigma}_{M_m}^{+}(\hat{\sigma}_{M_m}^{-})$  defines the raising (lowering) operator for $m=\{1,2\}$ respectively, such that $\hat{\sigma}_{M_m}^{+}\ket{0}_{M_m}=\ket{1}_{M_m}$ and $\hat{\sigma}_{M_m}^{-}\ket{1}_{M_m}=\ket{0}_{M_m}$. The qubits $M_m$ of the QATM are described by the thermal Gibbs states in the Boltzmann distribution ($k_B=1$) as:  
	\begin{equation}
		\hat{\rho}_{M_m}=\frac{1}{Z_{M_m}}e^{-\beta_{m}\hat{H}_{M_m}},\quad \hbox{for $m=\{1,2\}$},
	\end{equation}
	with $Z_{M_m}=Tr\big(e^{-\beta_{m}\hat{H}_{M_m}}\big)$. For the assumptions: $\omega_{M_1}<\omega_{M_2}$ and $\beta_{1}=:\frac{1}{T_1}\geq \beta_{2}=:\frac{1}{T_2}$, where $T_m$ is the temperature, the QATM acts as a single body noted $M_{12}$ with the corresponding spacing energy $\omega_{M_{12}}=\omega_{M_2}-\omega_{M_1}$, coupled to a virtual reservoir $R_v$ at the following virtual temperature $\mathcal{T}_v$ \cite{MODEL1}. Physically, the virtual reservoir is an effective heat bath composed by the thermal reservoirs $R_1$ and $R_2$ \cite{intro10}. It acts as a real reservoir for a specific energy transition in a quantum system, even though it does not exist as a physical entity. Its temperature, namely virtual temperature $\mathcal{T}_v$, can be higher, lower, or even negative compared to the temperatures of the real reservoirs $R_1$ and $R_2$. Indeed, the virtual temperature is expressed as follows: 
	\begin{equation}
		\mathcal{T}_{v} = \frac{\omega_{{M_{12}}}}{\omega_{M_2}\beta_{2} - \omega_{M_1}\beta_{1}}.
	\end{equation} 
	$\mathcal{T}_{v}$ is not the average temperature between the temperatures $T_1$ and $T_2$, while in the equilibrium case, one can write $T_1=T_2=\mathcal{T}_{v}$. The virtual temperature can also determine the type of QATM, such as the Heat pump, refrigerator, engine, etc \cite{MODEL1}. \\
	
	Let's introduce the Hamiltonian that describes, on one hand, the interaction between the field $f$ and the quantum charger, and on the other hand, the interaction between the quantum charger and the quantum battery. Indeed, it is expressed as follows:
	\begin{equation}\label{HCB}
			\hat{H}_{CB}(t)=\hat{H}_{CF}(t)+\hat{H}_{B}+\hat{H}_{C-B},
	\end{equation}
	where
		\begin{eqnarray}
			\hat{H}_{CF}(t)&=&\hat{H}_{C}+\Delta \hat{H}_{F}(t),\nonumber\\
			\hat{H}_{C-B}&=&k(\hat{\sigma}_{C}^{+}\hat{\sigma}_{B}^{-}+h.c),\nonumber\\
			\hat{H}_{n}&=&\omega_{n}\hat{\sigma}_{n}^{-}\hat{\sigma}_{n}^{+}~~(n=\{C,B\}),
	\end{eqnarray}
	where $\hat{H}_{CF}(t)$ is the charger-field Hamiltonian, and $\hat{H}_{C-B}$ denotes the interaction Hamiltonian between charger and battery. While $\hat{H}_{n}$, $n = \{C, B\}$, are the free Hamiltonians of the charger and battery, respectively. $\omega_n$ defines the energy level spacing's of the quantum charger $C$ and quantum battery $B$. The raising and lowering operators of the charger and battery satisfy $\hat{\sigma}_{n}^{+} \ket{0}_{n} = \ket{1}_{n}$ and $\hat{\sigma}_{n}^{-} \ket{1}_{n} = \ket{0}_{n}$.\\
	
	Besides, $\Delta \hat{H}_{F}(t)$ is the Hamiltonian of the field $f$, represented as the coherence driving of the quantum charger. In quantum thermodynamics, the coherence driving is described as the external work that can be given to a charger. Also, the role of the field is to preserve the energy of the quantum charger during the charging process, which can boost the energy transferred between the charger and the battery. It is not only an external resource of energy for the quantum charger, but can also improve the extractable work from the quantum battery \cite{intro11,intro12,intro13}:
	\begin{equation}
		\Delta\hat{H}_{F}(t)=f\Big(e^{-i\omega_{C}t}\hat{\sigma}_{C}^{+}+e^{+i\omega_{C}t}\hat{\sigma}_{C}^{-}\Big).
	\end{equation}
	The interaction between the QATM and quantum charger is based on the exchange of excitation between $M_{12}-C$ and its bias, as from $\ket{0_{M_1}1_{M_2}0_{C}}=\ket{1_{M_{12}}0_{C}}$ to $\ket{1_{M_1}0_{M_2}1_{C}}=\ket{0_{M_{12}}1_{C}}$. $\ket{0_{M_1}1_{M_2}}=\ket{1_{M_{12}}}$ and $\ket{1_{M_1}0_{M_2}}=\ket{0_{M_{12}}}$ prove that the QATM acts as a single body with their excited and ground states, $\ket{1_{M_{12}}}$ and $\ket{0_{M_{12}}}$ \cite{intro10a,MODEL1,ModelAch1,ModelAch2}.\\
	
	To maintain the conservation of energy between $M_{12}-C-B$, let’s consider the resonance case, $\omega_{M_{12}}=\omega_{C}=\omega_{B}$. The interaction between $M_{12}$ and the charger (C) is represented by the Hamiltonian,
	\begin{equation}
		\hat{H}_{M_{12}-C}=g\Big(\ket{1_{M_{12}}0_{C}}\bra{0_{M_{12}}1_{C}}+h.c\Big).
	\end{equation}
	The interaction between the reservoirs $R_1-R_2$ and the thermal machine qubits $M_1-M_2$, is defined by the interaction Hamiltonian $\hat{H}_{R-M_{12}}$,
	\begin{equation}
		\hat{H}_{R-M_{12}}=\sum_{m=1}^{2}\sum_k g_{mk}  \Big( \hat{\sigma}_{M_m}^{+} \otimes \hat{a}_{mk} + \hat{\sigma}_{M_m}^{-} \otimes \hat{a}_{mk}^{\dagger} \Big).
	\end{equation}
     The interaction between the quantum charger, battery, and  Markovian reservoirs is mediated by the presence of QATM, meaning that the QATM serves as a buffer or intermediary (filter) between the charger–battery system and the reservoirs, shaping how reservoir noise and decoherence influence the charger–battery dynamics.
	 When $M_{12}$ is reacted as a single body coupled to a virtual reservoir noted as $R_v$ at an equilibrium virtual temperature $\mathcal{T}_{v}$, the composite system $R_{v}-M_{12}$ is represented as a structured global bath $R_{v}M_{12}$ coupled to the charger and battery.  To evolve the dynamics of the system, let's consider the initial state of the composite system QATM-C-B as  
	\begin{equation}
		\hat{\rho}(0)=\hat{\rho}_{M_{12}}(0)\otimes\hat{\rho}_{C}(0)\otimes\hat{\rho}_{B}(0),
	\end{equation}
	$\hat{\rho}_{M_{12}}(0)$, $\hat{\rho}_{C}(0)$ and $\hat{\rho}_{B}(0)$ are the initial states of QATM, charger and battery. The reservoirs $R_1$ and $R_2$ are weakly coupled to their corresponding qubits $M_1$ and $M_2$. In this case, the local standard Born-Markov master equation in Lindblad form is \cite{MODEL2, MODEL3, MODEL4}
	\begin{equation}\label{MASTER_EQUATION}
		\frac{d}{d t} \hat{\rho}(t)=-i\left[\sum_{n=C,B}\hat{H}_{n}+\hat{H}_{M_{12}},\hat{\rho}(t)\right]+\varphi(t) \mathcal{L}(t)[\hat{\rho}(t)],
	\end{equation}
	where
	\begin{align}
		\mathcal{L}(t)[\hat{\rho}(t)]=&-i\left[\Delta \hat{H}_F(t)+\hat{H}_{C-B}+H_{M_{12}-C}, \hat{\rho}(t)\right]\nonumber\\
		&+\sum_{m=1}^{2}\mathcal{D}^{[T_{m}]}[\hat{\rho}(t)],\nonumber\\
		\mathcal{D}^{[T_{m}]}[\hat{\rho}(t)]=&\gamma_{m}(\bar{n}_{m}(T_{m},\omega_{M_m})+1)\mathcal{D}^{[\hat{\sigma}_{M_{m}}^{-}]}[\hat{\rho}(t)]+\nonumber\\
		&\gamma_{m}\bar{n}_{m}(T_{m},\omega_{M_m})\mathcal{D}^{[\hat{\sigma}_{M_{m}}^{+}]}[\hat{\rho}(t)]\nonumber,\\
		\mathcal{D}^{[\hat{\sigma}_{M_{m}}^{\pm}]}[\hat{\rho}(t)]&=\hat{\sigma}_{M_{m}}^{\pm}\hat{\rho}(t)\hat{\sigma}_{M_{m}}^{\mp}-\frac{1}{2}\{\hat{\sigma}_{M_{m}}^{\mp}\hat{\sigma}_{M_{m}}^{\pm},\hat{\rho}(t)\},\nonumber
	\end{align}
	$[.,.]$ denotes the usual commutator and $\gamma_{m}$ is the decoherence parameter; fixed the timescale of the dissipation process. The first term on the right-hand side of Eq.(\ref{MASTER_EQUATION}) describes the free evolution of the $\hat{\rho}(t)$ over time. However, the second term gives rise to the dissipation term due to the interaction with the environment, which is composed of QATM, $R_1$ and $R_2$. The function $\varphi(t)$ in Eq.(\ref{MASTER_EQUATION}) is a dimensionless function, which is used to run “on/off” the interaction between the reservoirs $R_1$ and $R_2$ with the thermal machine qubits $M_1$ and $M_2$. Indeed, $\varphi(t)=1$ for $t\in[0,\tau]$, while $\varphi(t)=0$ elsewhere, where $\tau$ is the interaction time. The average number of bosons in each reservoir $R_m$ ($m = \{1,2\}$) is given as:
	\begin{equation}
		\bar{n}(T_m,\omega_{M_m})=\frac{1}{\exp(\frac{\omega_{M_m}}{T_m})-1}.
	\end{equation}
	Since the local master equation (\ref{MASTER_EQUATION}) is derived  under Born-Markov approximation, the conditions on the coupling parameters are given as $(g, k, \gamma_{m}) \ll (\omega_{M_2}, \omega_{M_1})$ for $m = \{1,2\}$ since we work in the weak-coupling limit. In our simulations, we used the parameters $g/\omega_{M_2} \leq 0.1$, $k/\omega_{M_2} \leq 0.1$, and $\gamma_{m}/\omega_{M_2} \leq 0.1$.  
	These limits are consistent with the simulation of the experimental framework of superconducting qubits, where the energy spacing is considered in the GHz range. In contrast, the couplings are considered in the MHz range \cite{sup1,sup2,sup3}.
	%---------------------------------------------------------------------------------
	\subsection{Dynamical effects of QATM and the coherence driving on charger-battery system}\label{sec:Dynamical_effect_of_QATM} 
	
	In the time interval $t\in[0,\tau]$, the interaction between the subsystems $\hat{H}_{M_{12}-C}$ and $\hat{H}_{C-B}$ without coherence driving $\Delta\hat{H}_{T}(F)$ represents a conservative quantities during time, such that
	\begin{align}
		\left[\sum_{n=C}^{B}\hat{H}_{n}+\hat{H}_{M_{12}},\hat{H}_{M_{12}-C}\right]&=0,\\
		\left[\sum_{n=C}^{B}\hat{H}_{n}+\hat{H}_{M_{12}},\hat{H}_{C-B}\right]&=0,\\
		\left[\sum_{n=C}^{B}\hat{H}_{n}+\hat{H}_{M_{12}},\Delta\hat{H}_{F}(t)\right]&=\left[\hat{H}_{C},\Delta\hat{H}_{F}(t)\right]\nonumber\\
		&=f\omega_{C}(e^{i\omega_{C}t}\hat{\sigma}_{C}^{-}\nonumber\\&+e^{-i\omega_{C}t}\hat{\sigma}_{C}^{+})\label{Coherence_driving}.
	\end{align}
	The QATM, together with the quantum charger and battery, conserved their energies during their mutual interaction in the presence ($f \ne 0$) and also in the absence ($f = 0$) of coherence driving. This implies the QATM is operated independently, preserving the thermal machine's autonomy. In the numerical simulation, we set the energy scale $\omega_{M_2} = 10$ and set the different parameters $\omega_{M_1} = 0.2\omega_{M_2}$, $\gamma_1 = \gamma_2 = 0.02\omega_{M_2}$, $k = 0.03\omega_{M_2}$, $T_2 = 3\omega_{M_2}$, $T_1=0.1T_2$, These parameters are consistent with the weak coupling limits ($g/\omega_{M_2} \leq 0.1$, $k/\omega_{M_2} \leq 0.1$, and $\gamma_{m}/\omega_{M_2} \leq 0.1$), which align with the experimental framework of superconducting qubits~\cite{sup1,sup2,sup3}. Moreover, the initial state of the quantum charger-battery is written as
	\begin{equation}
		\hat{\rho}_{CB}(0) = \ket{1_{C}0_{B}}\bra{1_{C}0_{B}},
	\end{equation}
	with the initial energies for charger and battery are $E_{C}(0)/\omega_{C} = 1$ and $E_{B}(0)/\omega_{B} = 0$, respectively.
	%---------------------------------------------------------------------------------
	%---------------------------------------------------------------------------------	
	\subsubsection{Informations backflow and non-Markovian effects in subsystems interactions}\label{sec:NONMARKOVIANITY_MESURE}
	
	Information backflow between the subsystems (charger, battery and QATM) allows the presence of non-Markovianity, meaning physically there is a memory effect during the time evolution \cite{NONMARKOVIANITY1}. The derivative of the trace distance $\sigma_{n}(t)$ is used to quantify non-Markovianity. Indeed, it is given in terms of any two distinct states, namely $\hat{\rho}_{n}^{\alpha}(t)$ and $\hat{\rho}_{n}^{\beta}(t)$, with $n \in \{C, B, M_{12}\}$ as: 
	\begin{equation}
		\sigma_{n}(t)=\frac{1}{2}\frac{d}{dt}||\hat{\rho}_{n}^{\alpha}(t)-\hat{\rho}_{n}^{\beta}(t)||,
	\end{equation}
	where $||.|| = \text{Tr}\big(\sqrt{.^{\dagger}.}\big)$. For $\sigma_{n}(t)>0$, the subsystem $n = {M_{12}, C, B}$ loses its information to the environment. However, if it is negative, namely  $\sigma_{n}(t)>0$, there is a feedback of information from the environment to the system. If $\sigma_{n}(t)$ vanishes, then there is no change in information \cite{NONMARKOVIANITY2}. Concerning the quantum battery, we calculate the derivative of the trace distance between two initial states, $\hat{\rho}_{B}^{\alpha}(0) = \ket{0}_{B}\bra{0}$ and $\hat{\rho}_{B}^{\beta}(0) = \ket{1}_{B}\bra{1}$. For the quantum charger, we take as the initial states $\hat{\rho}_{C}^{\alpha}(0) = \hat{\rho}_{C}^{\beta}(0) = \ket{1}_{C}\bra{1}$, and for the  $M_{12}$, we used as the initial states $\hat{\rho}_{M_{12}}^{\alpha}(0) = \hat{\rho}_{M_{12}}^{\beta}(0) = \hat{\rho}_{M_{1}}(0) \otimes \hat{\rho}_{M_{2}}(0)$.\\
	
	%%%%%%%%%%%%%%%%%%%%%%%%%%%%%%%%FIG2%%%%%%%%%%%%%%%%%%%%%%%%%%%%%%%%%%%%%%%%%%%%%%%%%%%%%%%%%%%%%%%%%%%%%%%%%%%%%%%%%%%%%%
	\begin{figure*}[ht!]
		\subfloat[\label{NM_without_field}]{
			\includegraphics[width=1
			\columnwidth]{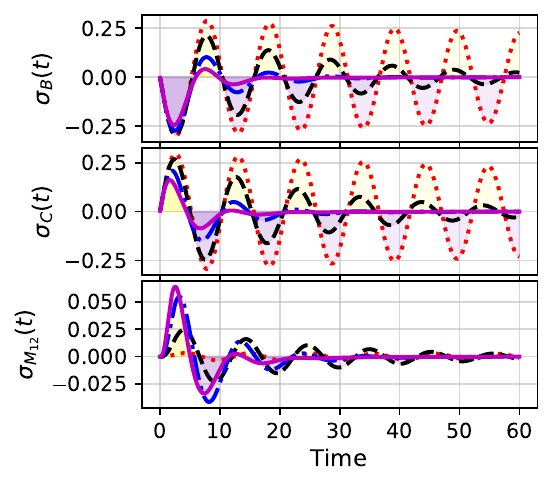}
		} \hfill
		\subfloat[\label{NM_with_field}]{
			\includegraphics[width=1
			\columnwidth]{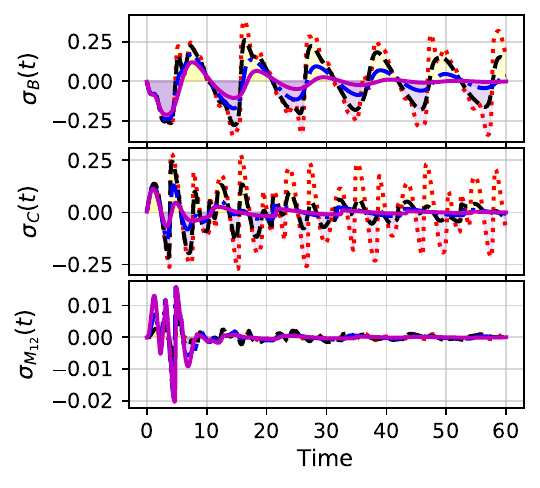}	
		}
		\caption{Plots of the trace distance derivative over time $\sigma_{B}(t)$, $\sigma_{C}(t)$, and $\sigma_{M_{12}}(t)$ of the quantum battery (B), quantum charger (C), and the QATM $M_{12}$, respectively, where  $g = 0.01\omega_{M_2}$, $0.03\omega_{M_2}$, $0.06\omega_{M_2}$, and $0.09\omega_{M_2}$ ('red, dotted line), (black, dashed line), (blue, dashed-dotted line), and (magenta, solid line), respectively. Figs (\ref{NM_without_field}) and (\ref{NM_with_field}) represent the cases of $f = 0$ and $f =0.1 \omega_{C}$. For $\sigma_{n}(t) > 0$ (yellow region), the subsystem $n = \{M_{12}, C, B\}$ loses information to its environment. If $\sigma_{n}(t) < 0$ (purple region), it gains information from the environment. } \label{NM}	
	\end{figure*}
	%%%%%%%%%%%%%%%%%%%%%%%%%%%%%%%%%%%%%%%%%%%%%%%%%%%%%%%%%%%%%%%%%%%%%%%%%%%%%%%%%%%%%%%%%%%%%%%%%%%%%%%%%%%%%%%%%%%%%%
	In Fig.(\ref{NM}), the backflow of information between the subsystems over time is investigated. In Fig.(\ref{NM_without_field}), the evolution of the derivative of trace distance without coherence driving on the quantum charger ($f = 0$) is considered. In this case, we observe in all subsystems a non-monotonic evolution of the trace distance derivative, $\sigma_{B}(t)$, $\sigma_{C}(t)$, and $\sigma_{M_{12}}(t)$. Moreover, an exchange of information between the charger, battery, and QATM is observed when $\sigma_{B}(t) < 0$, while $\sigma_{C}(t) > 0$ and $\sigma_{M_{12}}(t) > 0$, which means that the flow of information is manifested from the battery to the charger-QATM system. Conversely, when $\sigma_{B}(t)$ is positive, $\sigma_{C}(t)$ and $\sigma_{M_{12}}(t)$ are negative, meaning that the information flow is manifested from the charger-QATM system to the battery. These results prove the existence of a memory effect between the subsystems. From Fig.(\ref{NM_with_field}), we manipulate the backflow of information in the presence of coherence driving ($f = 0.1\omega_{C}$) on the quantum charger over time. One can see a non-monotonic behavior of the trace distance derivative over time for all subsystems, supporting the existence of memory effects in each subsystem. The degree of non-Markovianity for the quantum battery and the quantum charger increases when $f = 0.1\omega_{C}$, compared to the opposite case, $f = 0$. Hence, a decrease in the degree of non-Markovianity of QATM is observed.\\
	
	To understand the origin of this backflow information between the subsystems, we will quantify the information exchanged using conventional mutual information. In the next section, we will examine the mutual information between the charger and battery, as well as between QATM and the charger-battery system.
	
	%---------------------------------------------------------------------------------	
	\subsubsection{Amount of information in subsystems interactions} \label{sec:amount_MESURE}
	
	The amount of information between the subsystems over time gives an idea of the transfer of information. It describes how the memory effects arise from the interaction between the subsystems due to the exchange of correlations over time~\cite{NONMARKOVIANITY3,NONMARKOVIANITY4}. To achieve this goal, we investigate the conventional mutual information~\cite{NONMARKOVIANITY5} to quantify the mutual information between the quantum charger and the quantum battery, $I_{CB}(t)$. Moreover, we shall examine the mutual information between the QATM and the composite charger-battery system, that is, $I_{M_{12}CB}(t)$,
	\begin{eqnarray}\label{MIn}
		I_{M_{12}CB}(t)&=&S(\hat{\rho}_{M_{12}}(t))+S(\hat{\rho}_{C}(t))+S(\hat{\rho}_{B}(t))-S(\hat{\rho}(t)),\nonumber\\
		I_{CB}(t)&=&S(\hat{\rho}_{C}(t))+S(\hat{\rho}_{B}(t))-S(\hat{\rho}_{CB}(t)),
	\end{eqnarray}
	where $S(\hat{\rho}_{n}(t)) = -\mathrm{Tr}\big(\hat{\rho}_{n}(t) \log(\hat{\rho}_{n}(t))\big)$ is the Von-Neumann entropy. $\hat{\rho}_{CB}(t) = \mathrm{Tr}_{M_{12}}\big(\hat{\rho}(t)\big)$ is the reduced state of the composite charger-battery system.

	%%%%%%%%%%%%%%%%%%%%%%%%%%%%%%%%FIG3%%%%%%%%%%%%%%%%%%%%%%%%%%%%%%%%%%%%%%%%%%%%%%%%%%%%%%%%%%%%%%%%%%%%%%%%%%%%%%%%%%%%%%
	\begin{figure*}[ht!]
		\centering		
		\subfloat[\label{MI_without_field}]{
			\includegraphics[width=1
			\columnwidth]{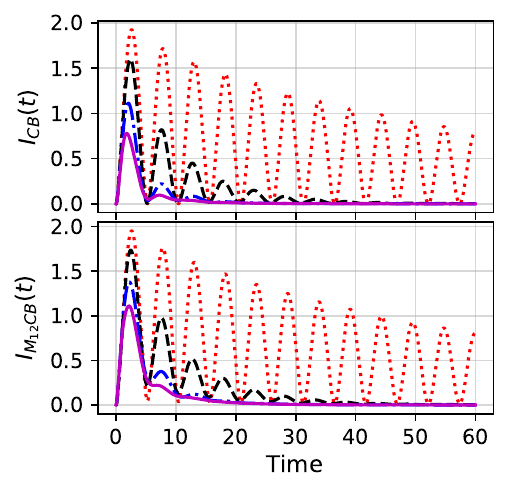}
		}\hfill
		\subfloat[\label{MI_with_field}]{
			\includegraphics[width=1
			\columnwidth]{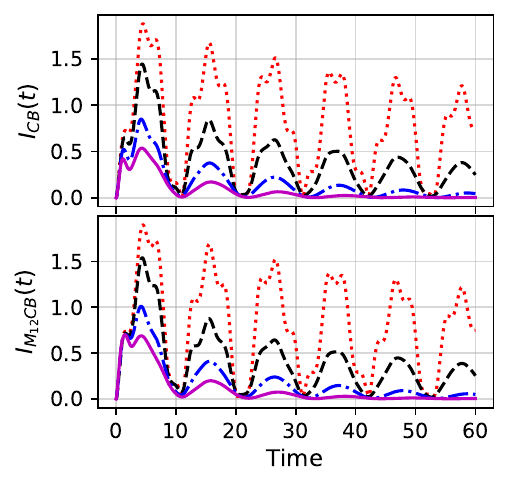}	
		}
		\caption{Plots of the conventional mutual information over time, $I_{CB}(t)$ between the charger (C) and the battery (B), and $I_{M_{12}CB}(t)$ between the set of charger-battery (CB) and the QATM ($M_{12}$). For the line styles corresponding to $g = 0.01\omega_{M_2}$, $0.03\omega_{M_2}$, $0.06\omega_{M_2}$, and $0.09\omega_{M_2}$, we use the same convention as in Figure~\ref{NM}. Figures~(\ref{MI_without_field}) and~(\ref{MI_with_field}) represent the cases of $f = 0$ and $f = 0.1\omega_{C}$, respectively. } \label{MI}	
	\end{figure*}
	%%%%%%%%%%%%%%%%%%%%%%%%%%%%%%%%%%%%%%%%%%%%%%%%%%%%%%%%%%%%%%%%%%%%%%%%%%%%%%%%%%%%%%%%%%%%%%%%%%%%%%%%%%%%%%%%%%%%%%
	The amount of information using mutual information given in Eq.(\ref{MIn}) is presented in Fig.(\ref{MI}). There are two cases, the first one examines the mutual information of $C$ without the effect of coherence driving ($f = 0$), Fig. (\ref{MI_without_field}). The second, Fig. (\ref{MI_with_field}) reflects the amount of the same quantity but in the presence of coherence driving ($f = 0.1\omega_{C}$). For  $I_{CB}(0) = I_{M_{12}CB}(0) = 0$, meaning $C, B$, and the QATM are initially uncorrelated. The mutual information $I_{CB}(t)$ and $I_{M_{12}CB}(t)$ increase non-monotonically over time, indicating an exchange of correlations. This is the main reason behind the existence of memory effects associated with non-Markovian dynamics, as shown in Figs. (\ref{MI_without_field}) and (\ref{NM_with_field}). Therefore, the presence of coherence driving enhances the amount of mutual information, indicating that the external field positively contributes to the increase in the presence of non-Markovianity in our system.\\
	
	From Fig. (\ref{MI}), the amount of information between the charger and the battery, $I_{CB}(t)$, is smaller than the amount of information between $M_{12}$ and the charger-battery system, $I_{M_{12}CB}(t)$, such, 
	
	\begin{align}
		I_{M_{12}-(CB)}(t)&=I_{M_{12}CB}(t) - I_{CB}(t),\nonumber\\
		I_{M_{12}-(CB)}(t) &= S(\hat{\rho}_{M_{12}}(t)) + S(\hat{\rho}_{CB}(t)) - S(\hat{\rho}(t)) \geq 0,
	\end{align}
	where $I_{M_{12}-(CB)}(t)$ denotes the mutual information between the QATM $M_{12}$ and the composite subsystem $(CB)$, i.e., the charger–battery system. Physically, this represents the difference between the correlations created within the composite system $M_{12}-(CB)$ and those between $M_{12}$ and charger (battery) individually. This difference gives rise to a stronger memory effect between the charger and the battery. In fact, the QATM $M_{12}$ acts as a filter for the decoherence effects due to the diversity between global and local correlation exchanges among the QATM, charger, and battery. Consequently, the degree of non-Markovianity of the charger, $\sigma_{C}(t)$, and that of the battery, $\sigma_{B}(t)$, are higher than that of QATM, which explains the non-Markovian behaviors observed in Figs~(\ref{NM_without_field}) and~(\ref{NM_with_field}).
	\\

	%---------------------------------------------------------------------------------
	\section{Impact of QATM and coherence driving on charging process}\label{sec:Impact_of_QATM_Coherence-driving_on_chargig_process}
	
	Let's consider the initial state of the quantum charger: 
	\begin{equation}
		\hat{\rho}_{C}(0)=\frac{1}{2}\Big(\ket{0}_{C}\bra{0}+\ket{1}_{C}\bra{1}+\ket{0}_{C}\bra{1}+\ket{1}_{C}\bra{0}\Big),
	\end{equation}
	with the initial energy $\frac{E_{C}(0)}{\omega_{C}}=\frac{1}{2}$. The initial state of the quantum battery is $\hat{\rho}_{B}(0)=\ket{0}_{B}\bra{0}$, with $\frac{E_{B}(0)}{\omega_{B}}=0$.
	%---------------------------------------------------------------------------------
	\subsection{Transfer of energy and coherence }\label{sec:Energy_and_cohrence_transfer}
	
	In this section, we quantify the transfer of energy using the internal energy of each subsystem, and the coherence of the quantum charger (C) and quantum battery (B) using the relative entropy of coherence, known as the Kullback--Leibler (KL) coherence divergence. This quantity describes the variation in coherence between the charger and the battery over time~\cite{intro20, intro21}.
	
	\subsubsection{Internal energy transfer} 
	The internal energies of the quantum charger and quantum battery, $\Delta E_{C}(t)$ and $\Delta E_{B}(t)$, are given in the compact form:
	\begin{align}
		E_n(t) &= \mathrm{Tr}\big[\hat{H}_n\hat{\rho}_n(t)\big] \quad (n=\{C,B,\}),\nonumber\\
		\Delta E_{n}(t) &= E_n(t) - E_n(0),
	\end{align}
where $E_n(0)$ and $E_n(t)$ denote the initial and time-dependent energies. While, $\Delta E_{n}(t)$ is measured relatively to the initial state i.e., $\Delta E_n(t) = E_n(t) - E_n(0)$.
	 For QATM ($M_{12}$), we have:
	\begin{align}
		E_{M_{m}}(t) &= \mathrm{Tr}\big[\hat{H}_{M_{m}}\hat{\rho}_{M_m}(t)\big] \quad (m=\{1,2\}),\nonumber\\
		\Delta E_{M_{12}}(t) &= \sum_{m=1}^{2}\frac{1}{\omega_{M_m}} \left(E_{M_m}(t) - E_{M_m}(0)\right),
	\end{align}
	where $E_{M_{m}}(t)$ are the energies over time of qubits $M_1$ and $M_2$ of the QATM. $\Delta E_{M_{12}}(t)$ denotes the normalized internal energy over time of $M_{12}$.\\
	
	%%%%%%%%%%%%%%%%%%%%%%%%%%%%%%%%FIG4%%%%%%%%%%%%%%%%%%%%%%%%%%%%%%%%%%%%%%%%%%%%%%%%%%%%%%%%%%%%%%%%%%%%%%%%%%%%%%%%%%%%%
	\begin{figure*}[ht!]
		\centering		
		\subfloat[\label{En_witout_field}]{
			\includegraphics[width=1.01
			\columnwidth]{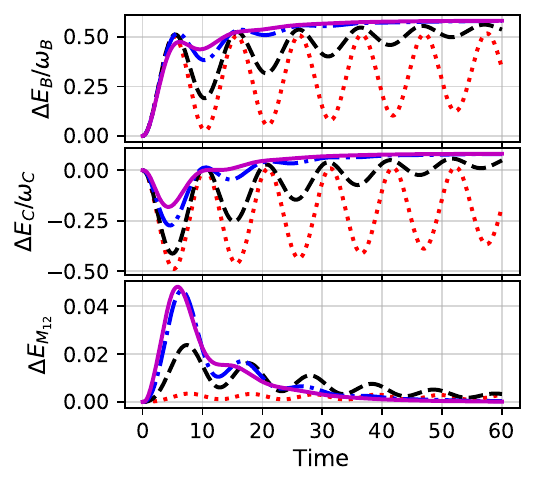}
		}\hfill
		\subfloat[\label{En_with_field}]{
			\includegraphics[width=1.01
			\columnwidth]{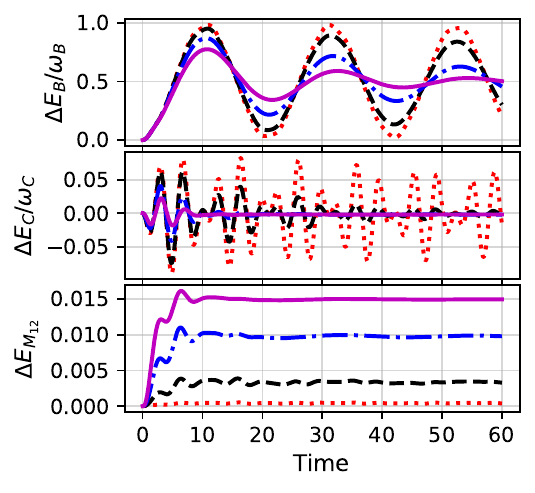}	
		}
		\caption{Plots of the normalized internal energies over time, $\frac{\Delta E_C(t)}{\omega_C}$, $\frac{\Delta E_B(t)}{\omega_B}$, and $\Delta E_{M_{12}}(t)$, of the quantum charger (C), quantum battery (B), and the QATM $M_{12}$, respectively.  For the line styles corresponding to  $g = 0.01\omega_{M_2}$, $0.03\omega_{M_2}$, $0.06\omega_{M_2}$, and $0.09\omega_{M_2}$, we use the same convention as in Figure~\ref{NM}. Figure~\ref{En_witout_field} represents the case of $f = 0$, while Figure~\ref{En_with_field} represents the case of $f = 0.1\omega_{C}$.} \label{EN}	
	\end{figure*}
	%%%%%%%%%%%%%%%%%%%%%%%%%%%%%%%%%%%%%%%%%%%%%%%%%%%%%%%%%%%%%%%%%%%%%%%%%%%%%%%%%%%%%%%%%%%%%%%%%%%%%%%%%%%%%%%%%%%%%%
	Fig.(\ref{EN}) displays the variation of the internal energies of $C$ and $B$ over time under the effect of QATM ($M_{12}$) and the coherence driving ($f$) into the quantum charger. Indeed, Fig. (\ref{En_witout_field}) shows the internal energy variations of quantum charger, quantum battery, and QATM in the absence of a field $f = 0$, meaning without the injection of the field into the quantum charger over time,  $\left[\sum_{n=C}^{B}\hat{H}_{n}+\hat{H}_{M_{12}},\Delta\hat{H}_{F}(t)\right]=0$. The internal energy of the quantum battery $\Delta E_{B}(t)$ vanishes and then increases over time, which mathematically means $E_B(t) > E_B(0)$. Besides, the internal energy of the quantum charger $\Delta E_{C}(t)$, it also initially vanishes and decreases over time ($E_C(0) > E_C(t)$). However, there is a slight increase in the energy of the QATM over time, denoted by $\Delta E_{M_{12}}(t)$.
	\\
	The transfer of energy from the quantum charger into the quantum battery is filtered through the reservoirs $R_1$ and $R_2$ via $M_{12}$. At times $t > 20$, we have $E_C(t>20) > E_C(0)$ indicating that the charger is also recharged from the QATM due to the biased transition from $\ket{0_{M_1}1_{M_2}0_{C}}$ to $\ket{1_{M_1}0_{M_2}1_{C}}$, because the thermal machine $M_{12}$ acts as a heat pump ($\mathcal{T} = -2.4 < T_1 < T_2$)~\cite{MODEL1}, according to an inversion of population in the quantum charger.\\
	
	Fig. (\ref{En_with_field}) depicts the internal energy variations of the quantum charger, quantum battery and QATM in the presence of the field $f = \omega_C$, corresponding to the injection of a field from a laser into the quantum charger. Moreover, the variation of the charger's internal energy is slight over time ($\Delta E_C(t) \to 0$), indicating a conservation of the charger's energy over time due to the injection of internal energy, as described in Eq.(\ref{Coherence_driving}). This reflects on the important role of energy transfer between the charger and battery.\\
	
\begin{figure*}[ht!]
		\centering		
		\subfloat[\label{po_witout_field}]{
			\includegraphics[width=1
			\columnwidth]{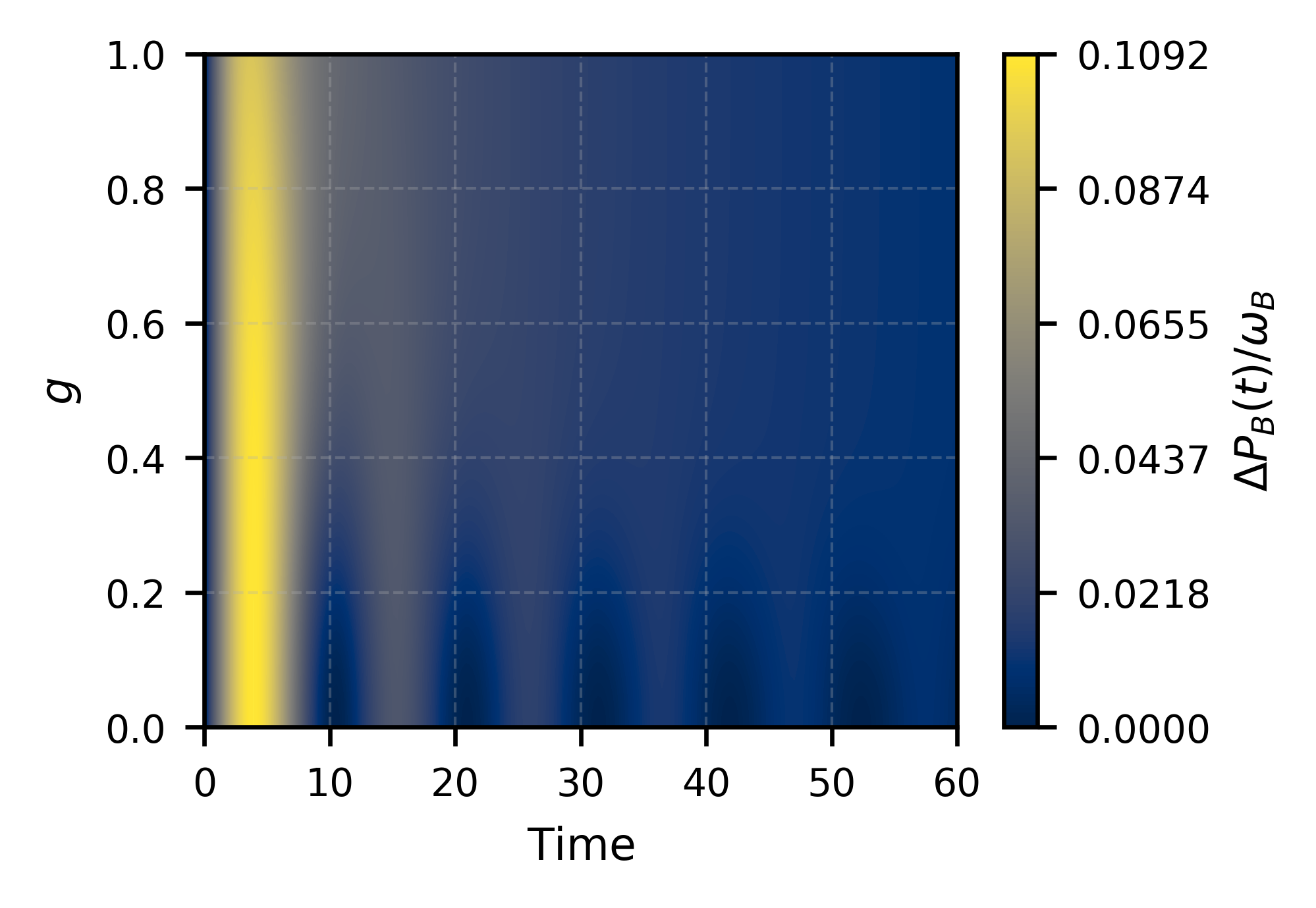}
		}
		\subfloat[\label{po_with_field}]{
			\includegraphics[width=1
			\columnwidth]{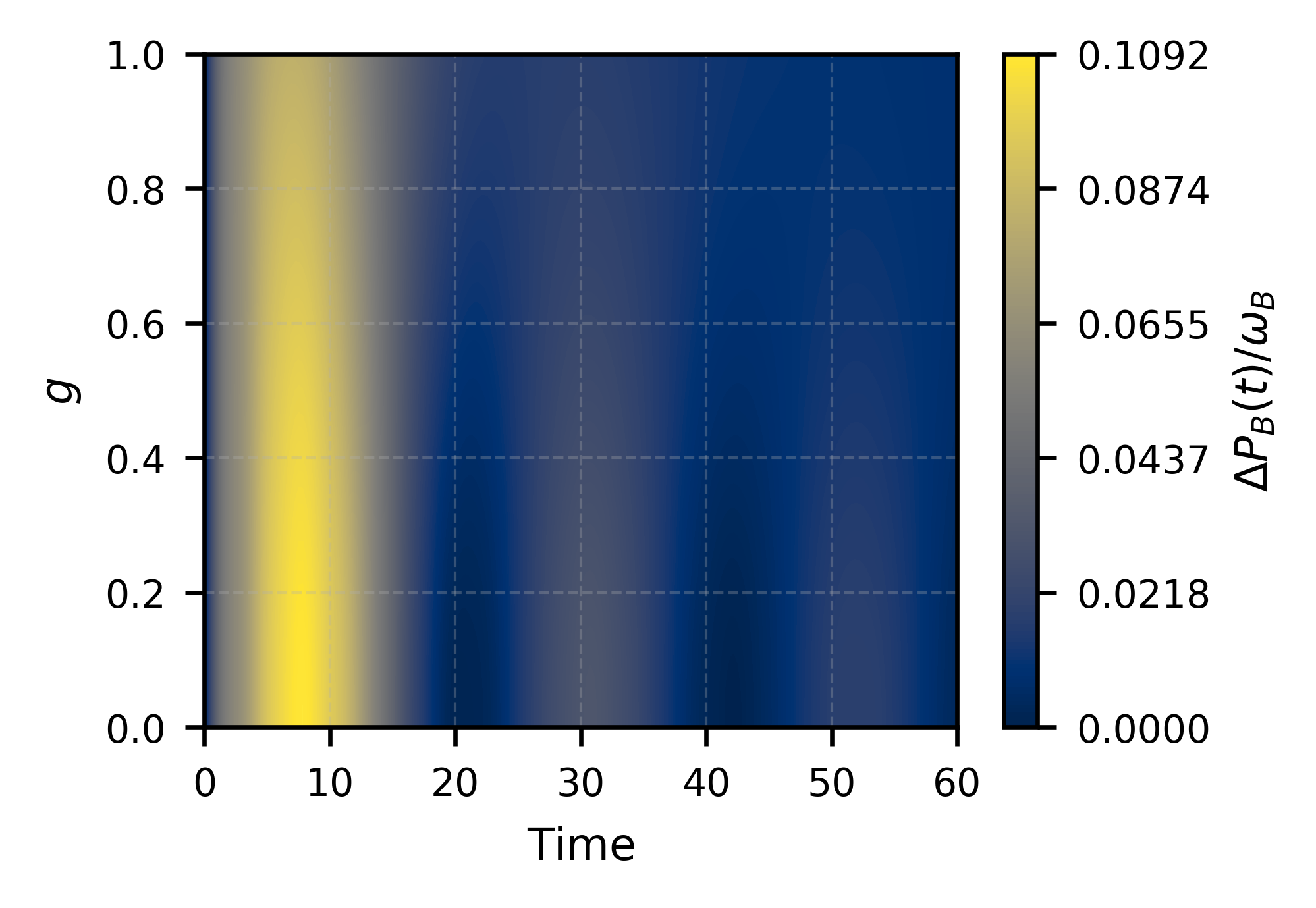}	
		}
		\caption{The density plot of the normalized power of charging for the quantum battery (B) over time and the coupling $g$, given by $\frac{\Delta P_{B}(t)}{\omega_{B}}$. Figure~\ref{po_witout_field} represents the case of $f = 0$, while Figure~\ref{po_with_field} represents the case of $f =0.1 \omega_{C}$. Here $g$ in unit of ($\omega_{M_2}/10$).
		} \label{po}	
	\end{figure*}

	To investigate how the coherence driving $f$ and the coupling $g$ between charger and QATM affect the quantum charging power $\Delta P_{B}(t)$ of the battery, we will analyze its dynamics under various considerations. Indeed, the quantum charging power $\Delta P_{B}(t)$ is defined as
	\begin{equation}
		\Delta P_{B}(t) = \frac{\Delta E_B (t)}{t},
	\end{equation}
	$\Delta E_B (t)$ denotes the energy of the quantum battery. In Fig. (\ref{po}), the density plot of $\Delta P_{B}(t)$ versus $t$ and $g$ is presented. The power of charging is plotted, in the case of the absence (presence) of coherence driving, $f = 0$ ($f = 0.1\omega_{C}$), on the charger, in Fig. (\ref{po_witout_field}) ( \ref{po_with_field}). The obtained results show that the charging power remains robust in the presence of coherence driving in the quantum charger comparing to the opposite case, which enhances the charging capacity of the quantum battery. In particular, the normalized charging power reaches a maximum value of approximately $\Delta P_B / \omega_B \approx 0.1$ around $t \sim 20$, which is nearly twice the maximum value obtained in the incoherent case ($f = 0$). This comparison emphasizes the coherence-assisted enhancement of the charging process. In the case without coherence driving, as shown in Fig. (\ref{po_witout_field}), the impact of the coupling $g$ is remarkable on the non-Markovianity because the coupling between the charger and the QATM basically destroys the non-Markovianity. Consequently, when the strength $g$ increases, then the variation of $\Delta P_{B}$ decreases monotonically over time.\\
	
	%%%%%%%%%%%%%%%%%%%%%%%%%%%%%%%%FIG5%%%%%%%%%%%%%%%%%%%%%%%%%%%%%%%%%%%%%%%%%%%%%%%%%%%%%%%%%%%%%%%%%%%%%%%%%%%%%%%%%%%%%
	%%%%%%%%%%%%%%%%%%%%%%%%%%%%%%%%%%%%%%%%%%%%%%%%%%%%%%%%%%%%%%%%%%%%%%%%%%%%%%%%%%%%%%%%%%%%%%%%%%%%%%%%%%%%%%%%%%%%%%
	
	%---------------------------------------------------------------------------------
	\subsubsection{Coherence transfer}\label{sec:coherence_transter}
	
	Let's quantify the coherence of the charger and the battery employing the relative entropy of coherence. The latter measure has been introduced to investigate the distance between a quantum state and its corresponding fully-dephased state. The relative entropies of coherence over time for the quantum charger and the quantum battery  are given as
	\begin{eqnarray}
		C(\hat{\rho}_{C}(t))&=&S(\hat{\rho}_{C}(t)||\tilde{\rho}_{C}(t)),\nonumber\\
		C(\hat{\rho}_{B}(t))&=&S(\hat{\rho}_{B}(t)||\tilde{\rho}_{B}(t)),
	\end{eqnarray}
	where, 
	$$S(\hat{\rho}_{n}(t)||\tilde{\rho}_{n}(t))= Tr_{n}\big(\hat{\rho}_{n}(t)\log\hat{\rho}_{n}(t) - \hat{\rho}_{n}(t)\log\tilde{\rho}_{n}(t)\big),$$ such that  
 $\tilde{\rho}_{n}(t)$ is the fully dephased state expressed as follows:  
		\begin{equation}
			\tilde{\rho}_{n}(t)=\sum_{j}\ket{j}_{n}\bra{j}_{n}\hat{\rho}_{n}(t)\ket{j}_{n}\bra{j}_{n},~~n=\{C,B\}. 
	\end{equation}
	The above equation shows that the density matrix of the quantum charger or quantum battery is a density matrix without the off-diagonal elements.\\
	
	%%%%%%%%%%%%%%%%%%%%%%%%%%%%%%%%FIG6%%%%%%%%%%%%%%%%%%%%%%%%%%%%%%%%%%%%%%%%%%%%%%%%%%%%%%%%%%%%%%%%%%%%%%%%%%%%%%%%%%%%%
	\begin{figure*}[ht!]
		\centering		
		\subfloat[\label{C_witout_field}]{
			\includegraphics[width=0.99
			\columnwidth]{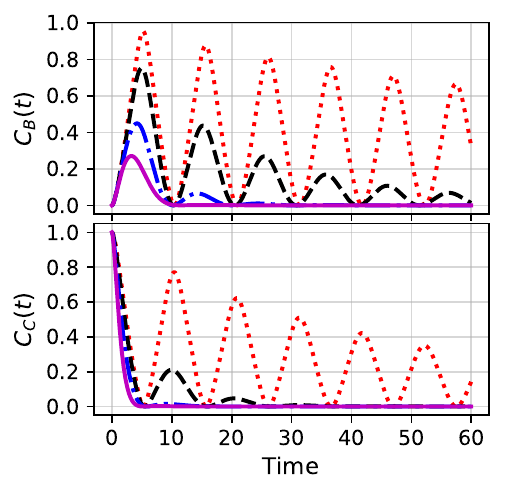}
		}
		\subfloat[\label{C_with_field}]{
			\includegraphics[width=0.99
			\columnwidth]{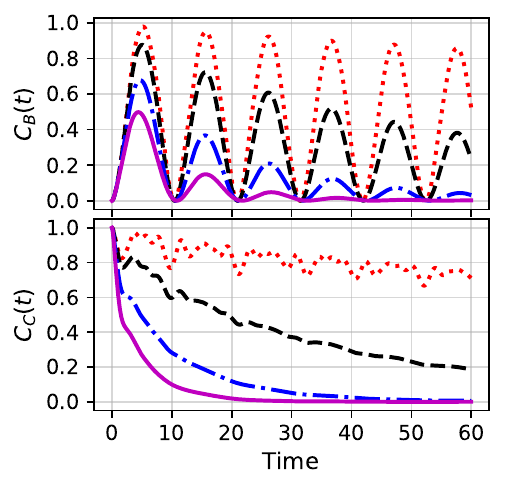}	
		}
		\caption{Plots of the relative entropy of coherence versus time, namely $C(\hat{\rho}_{C}(t))$ and $C(\hat{\rho}_{B}(t))$ for quantum charger (C) and quantum battery (B), respectively. The line styles correspond to $g = 0.01\omega_{M_2}$, $0.03\omega_{M_2}$, $0.06\omega_{M_2}$, and $0.09\omega_{M_2}$, following the same convention as in Fig~(\ref{NM}). Moreover, we set $f = 0, 0.1$ for  Figs~(\ref{C_witout_field}) and~(\ref{C_with_field} ), respectively	} \label{C}	
	\end{figure*}
	%%%%%%%%%%%%%%%%%%%%%%%%%%%%%%%%%%%%%%%%%%%%%%%%%%%%%%%%%%%%%%%%%%%%%%%%%%%%%%%%%%%%%%%%%%%%%%%%%%%%%%%%%%%%%%%%%%%%%%
	
	The relative entropy of coherence for the quantum charger and quantum battery is plotted, Fig. (\ref{C}), to manipulate the effect of QATM, charger interaction $g$, and coherence driving into the quantum charger. From Fig. (\ref{C_witout_field}), we represent the effect of the interaction between QATM and the quantum charger without the coherence driving. Hence, the coherence of the quantum charger decreases non-monotonically, and the quantum battery increases non-monotonically over time. Moreover, the maximal amplitude of the coherence of $B$ gives rise to a minimal coherence of $C$. This means that the non-monotonic evolution of the coherence of $C$ and $B$ is due to the non-Markovian dynamics and the correlation exchange between them. In Fig. (\ref{C_with_field}), the amount of coherence of $C$ and $B$ in the presence of coherence driving is shown. The evolution of $C_C(t)$ is monotonic over time. However, the dynamics of $C_B(t)$ is non-monotonic over time.\\
	
	%---------------------------------------------------------------------------------------------
	\subsection{Ergotropy of the quantum battery}\label{sec:ergotropy_of_quantum_battery}
	
	The quantum ergotropy quantifies the amount of maximal work that a quantum system can extract under the action of a unitary operation. In this context, quantum batteries are intriguing for their ability to describe the maximum amount of work that can be extracted. The ergotropy reflects the difference between quantum battery energy $E_{B}(t)$ and the energy of its passive state,
	\begin{eqnarray}\label{pass}
		E_{B_{P}}(t)&=&\mathrm{Tr}\big(\hat{H}_{B}\hat{\rho}_{B_P}(t)\big)\nonumber\\
		&=&\min_{\hat{U}_{B}}\mathrm{Tr}\big(\hat{H}_{B}\hat{U}_{B}\hat{\rho}_{B}(t)\hat{U}_{B}^{\dagger}\big).
	\end{eqnarray}
	$\hat{U}_{B}$ denotes the unitary evolution operator of the quantum battery, $\hat{\rho}_{B_P}$ is the passive state of the quantum battery. The corresponding Hamiltonian of the quantum battery is $\hat{H}_{B}=\sum_{j}\omega_{j}\ket{\omega_{j}}\bra{\omega_j}$, where $\omega_{j}$ are its eigenvalues ($\omega_1\leq\omega_2\leq\ldots$) and $\ket{\omega_{j}}$ define the corresponding eigenstates. The passive state $\hat{\rho}_{B_P}$ is defined as
	\begin{equation}
		\hat{\rho}_{B_P}(t)=\sum_{j}P^{B}_{j}(t)\ket{\omega_{j}}\bra{\omega_j},
	\end{equation}
	where $P^{B}_{j}(t)$ are the populations of $\hat{\rho}_{B}(t)$ reordered as $P^{B}_{1}(t)\geq P^{B}_{2}(t)\geq ...$, ensuring that higher population corresponds to lower eigenvalue. Hence, the quantum ergotropy is given as
	\begin{equation}
		\epsilon_{B}= E_{B}(t) - E_{B_{P}}(t).
	\end{equation}
	From Eq.(\ref{pass}), $\hat{U}_{B}\hat{\rho}_{B}(t)\hat{U}_{B}^{\dagger}$ is the state of the battery under any possible unitary operation $\hat{U}_{B}$. Therefore, the energy of the passive state of quantum battery corresponds to its minimum energy, that is, $\mathrm{Tr}\big(\hat{H}_{B}\hat{\rho}_{B_P}(t)\big)$.
	\\
	
	%%%%%%%%%%%%%%%%%%%%%%%%%%%Figure 7 %%%%%%%%%%%%%%%%%%%%%%%%%%%%%%%%%%%%%%%%%%%%%%%%%%%%%%
	\begin{figure*}[ht!]
		\centering		
		\subfloat[\label{Ergo_without_filed}]{
			\includegraphics[width=1
			\columnwidth]{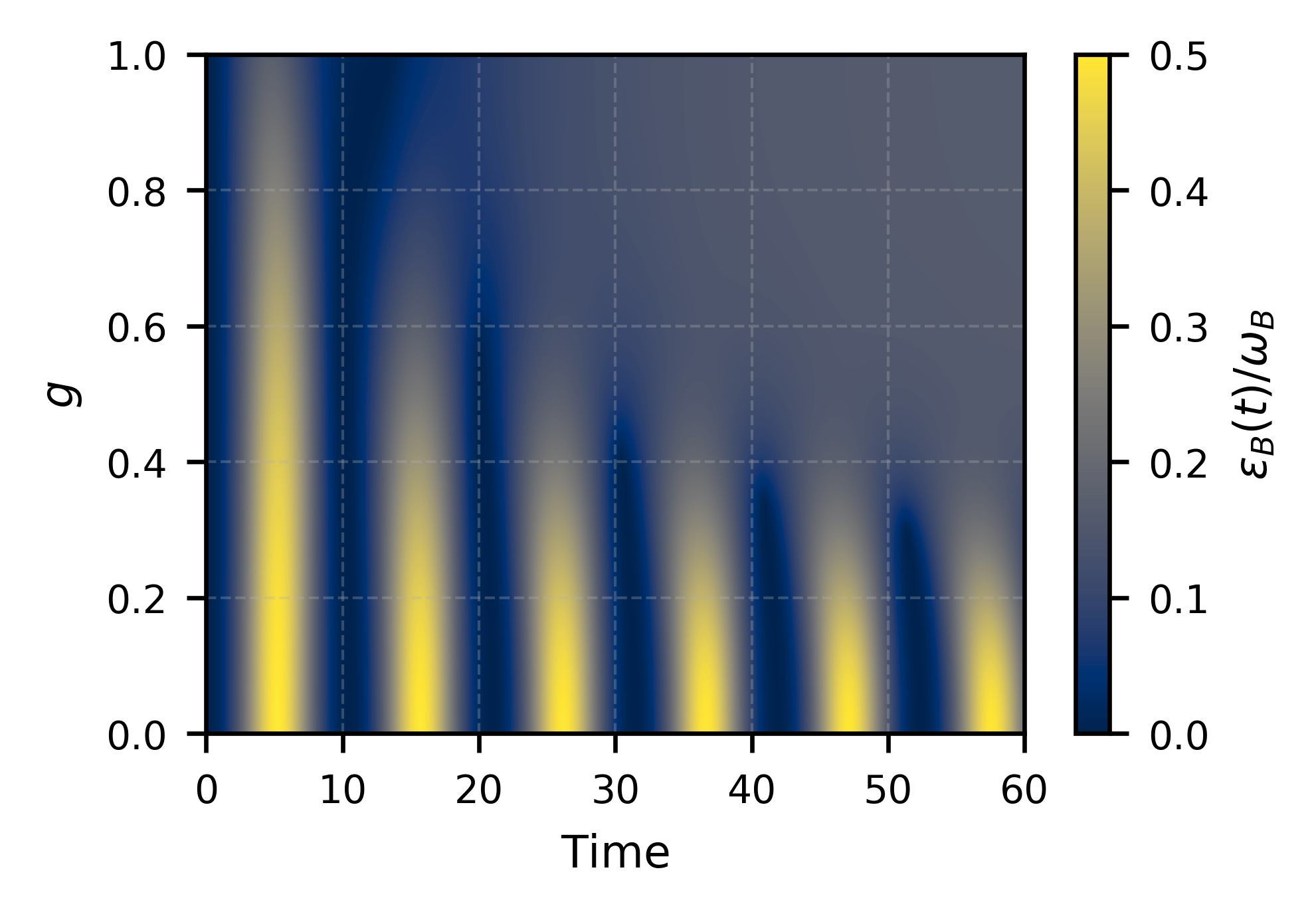}
		}
		\subfloat[\label{Ergo_with_filed}]{
			\includegraphics[width=1
			\columnwidth]{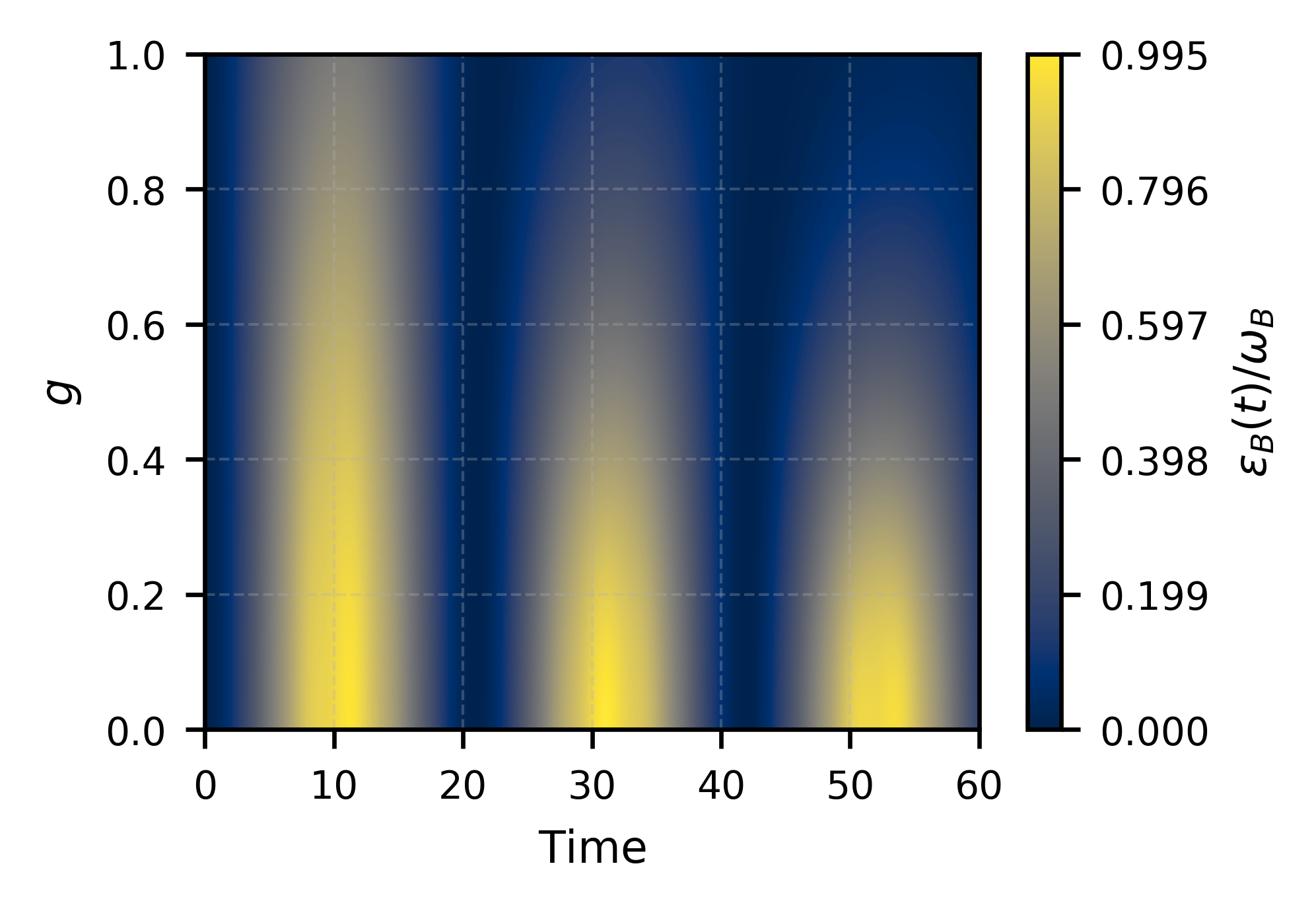}	
		}
		\caption{Density plot of the normalized ergotropy of the quantum battery (B) over time and the coupling $g$, given by $\frac{\epsilon_{B}(t)}{\omega_{B}}$. Figure~\ref{po_witout_field} represents the case of $f = 0$, while Figure~\ref{po_with_field} represents the case of $f =0.1
			\omega_{C}$. Here $g$ in unit of ($\omega_{M_2}/10$).
		} \label{Ergo}	
	\end{figure*}
	%%%%%%%%%%%%%%%%%%%%%%%%%%%%%%%%%%%%%%%%%%%%%%%%%%%%%%%%%%%%%%%%%%%%%%%%%%%%%%%%%%%%%%%%%%%%%%%%%%%%%%%%%%%%%%%%%%%%%%
	Fig. (\ref{Ergo}) depicts the impact of the coupling strength between $C$ and $M_{12}$, on the dynamics of ergotropy for different values of $g$ and $f$. Fig.(\ref{Ergo_without_filed}) presents the amount of ergotropy $\epsilon_{B}(t)$ in the absence of coherence driving on the quantum charger. However, Fig.(\ref{Ergo_with_filed}) displays the same quantity but in the presence of coherence driving. For any value of $g$, the ergotropy initially vanishes for the two plots. Effectively, the initial state of the quantum battery $\hat{\rho}_{B}(0)=\ket{0}_{B}\bra{0}$ is proportional to its passive state $\hat{\rho}_{B_P}(0)=\ket{0}_{B}\bra{0}$, where the Hamiltonian of the quantum battery is reformulated as $\hat{H}_{B}= 0\times \ket{0}_{B}\bra{0}+\omega_{B}\ket{1}_{B}\bra{1}$.\\
	
    The ergotropy of the quantum battery evolves non-monotonically over time for small values of $g$ and vice versa, as shown in Fig.~(\ref{Ergo_without_filed}). This behavior is due to the non-Markovian dynamics, since the coupling strength $g$ between charger and QATM influences the suppression of non-Markovian effects. As illustrated in Fig.~(\ref{C_witout_field}), robust amplitudes of the quantum battery coherence, $C_B(t)$, give rise to maximum values of the ergotropy $\epsilon_B(t)$ and vice-versa. Importantly, coherence driving leads to approximately a 40\% increase in the maximum ergotropy compared to the case without coherence driving. This result indicates that coherence storage enhances basically the maximal work that can be extracted from the quantum battery over time.\\
	
	Let's discuss physically the impact of coherence driving $f=0.1\omega_{C}$ on ergotropy given in Fig. (\ref{Ergo_with_filed}). The amplitude of quantum battery ergotropy increases compared to the case of  $f=0$. Its evolution over time for any value of $g$ is non-monotonic over time, due to the non-Markovian dynamics of the quantum battery. As discussed in \ref{sec:NONMARKOVIANITY_MESURE}, the presence of the field increases the non-Markovianity of the quantum battery and quantum charger. Then, the maximal extracted work from the quantum battery is obtained when the coherence of the quantum battery is maximized.
	
	\section{Impact of the non-Markovianity on charging process}\label{sec:Impact_the_nonMarkovianity_on_chargig_process}
	
	The existence of the memory effect in the proposed model is due to the non-Markovian dynamics between the quantum charger and quantum battery under the influence of decoherence filtering of reservoirs $R_1$ and $R_2$ via $M_{12}$. The QATM is used as a thermodynamic resource to enhance the non-Markovianity and the exchange of correlation between the charger and battery. Let's fix the coupling between the quantum charger and QATM for $g=0.03\omega_{M_2}$, meaning the decoherence effects of the reservoirs $R_1$ and $R_2$ on the composite system charger-battery are set. The impact of the coupling between charger and battery, $k$, is responsible for improving correlation and non-Markovianity between them. In the absence of coherence driving and according to Eq.(\ref{Coherence_driving}), the parameter $f$ operates independently of the interactions between charger-QATM or charger-battery. The initial state of the quantum charger and battery is defined as:  
	\begin{eqnarray}
		\hat{\rho}_{C}(0) &=& \frac{1}{2}(\ket{1}_{C}\bra{1}+\ket{0}_{C}\bra{0}+\ket{0}_{C}\bra{1}+\ket{1}_{C}\bra{0}),\nonumber\\
		\hat{\rho}_{B}(0) &=& \ket{0}_{B}\bra{0},
	\end{eqnarray}
	where the initial energies $E_{C}(0)/\omega_{C} = 1$ and $E_{B}(0)/\omega_{B} = 0$ for the charger and battery.\\
	
	%%%%%%%%%%%%%%%%%%%%%%%%%%%%%%%%% FIGURE 8 %%%%%%%%%%%%%%%%%%%%%%%%%%%%%%%%%%%%%%%%%%%%%%%%%%%%%%%%%%%%%%%%%%%%
	\begin{figure*}[htb]
		
		\subfloat[\label{subfig:TD}]{%
			\includegraphics[width=1\columnwidth]{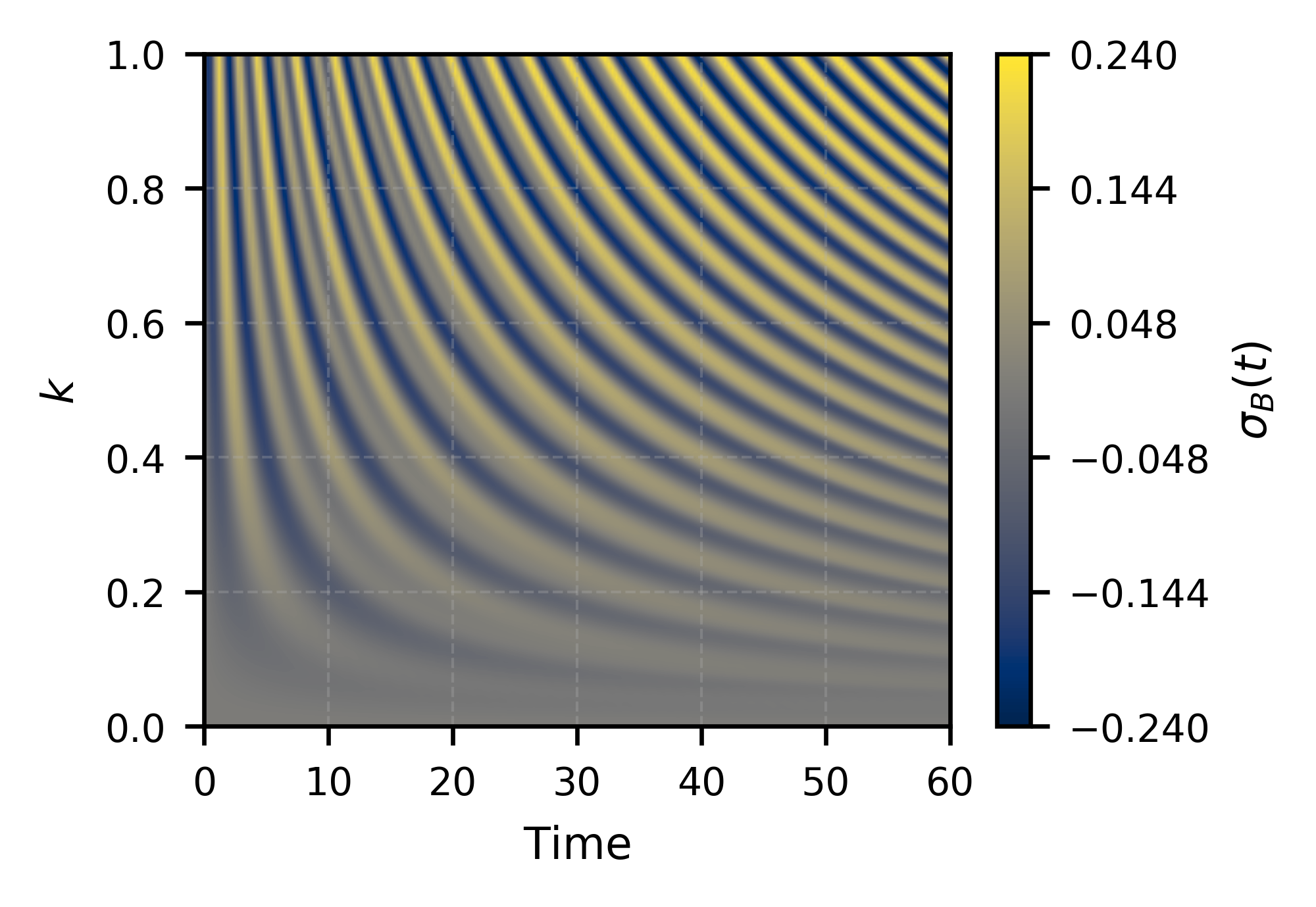}%
		}\hfill
		\subfloat[ \label{subfig:COB}]{%
			\includegraphics[width=1\columnwidth]{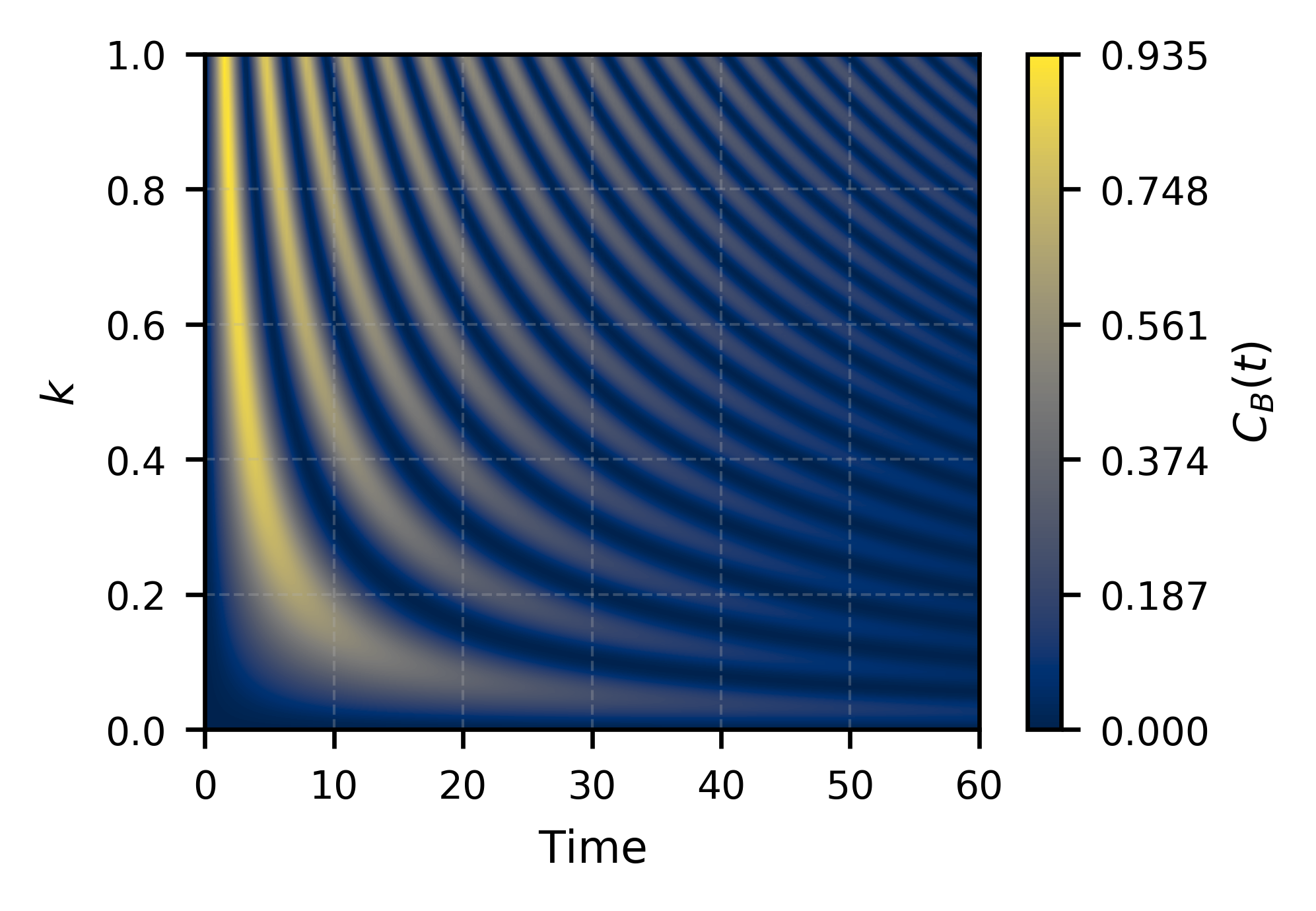}%
		}\hfill
		\subfloat[\label{subfig:Po}]{%
			\includegraphics[width=1\columnwidth]{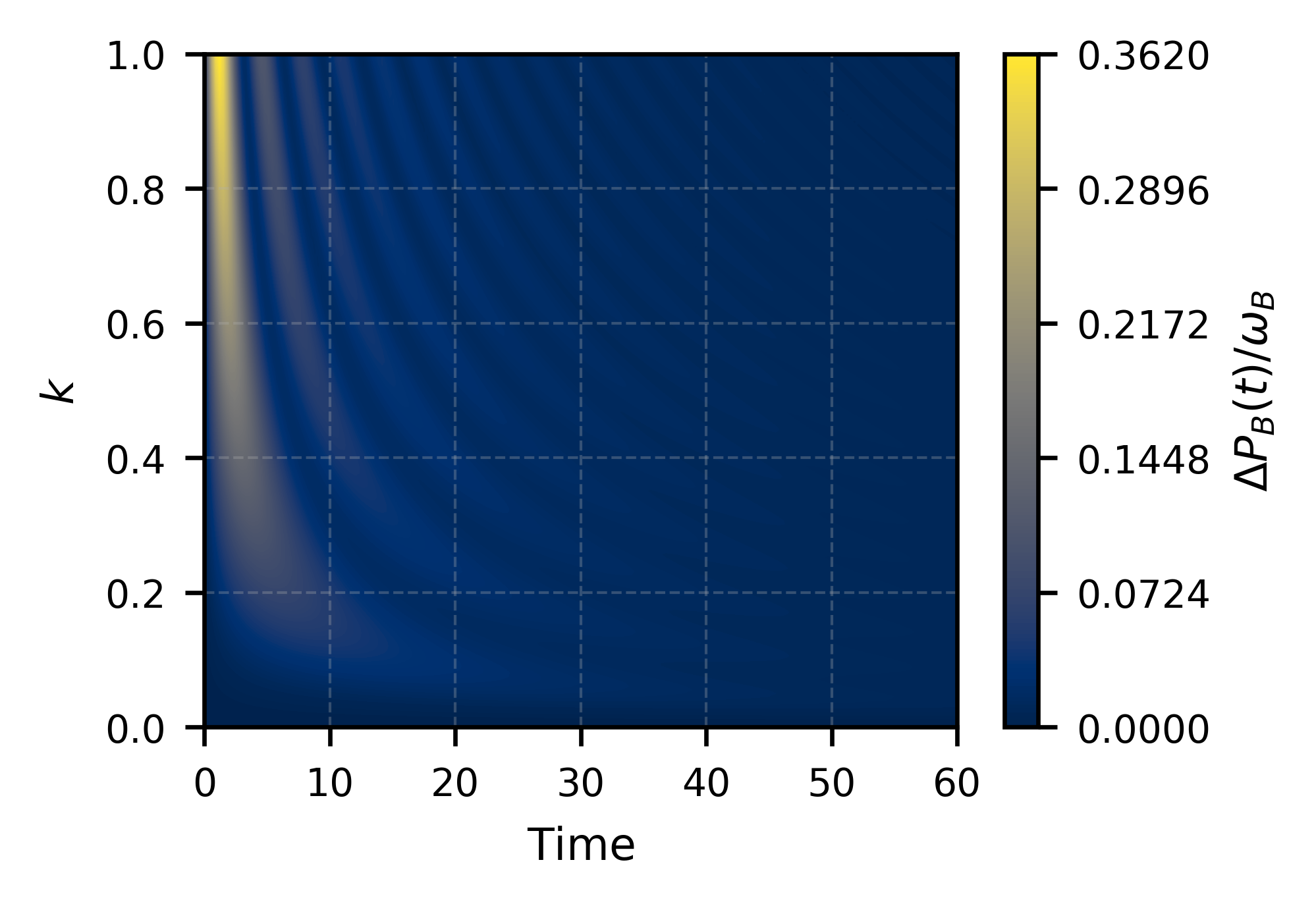}%
		}
		\hfill
		\subfloat[\label{subfig:ERGO}]{%
			\includegraphics[width=1\columnwidth]{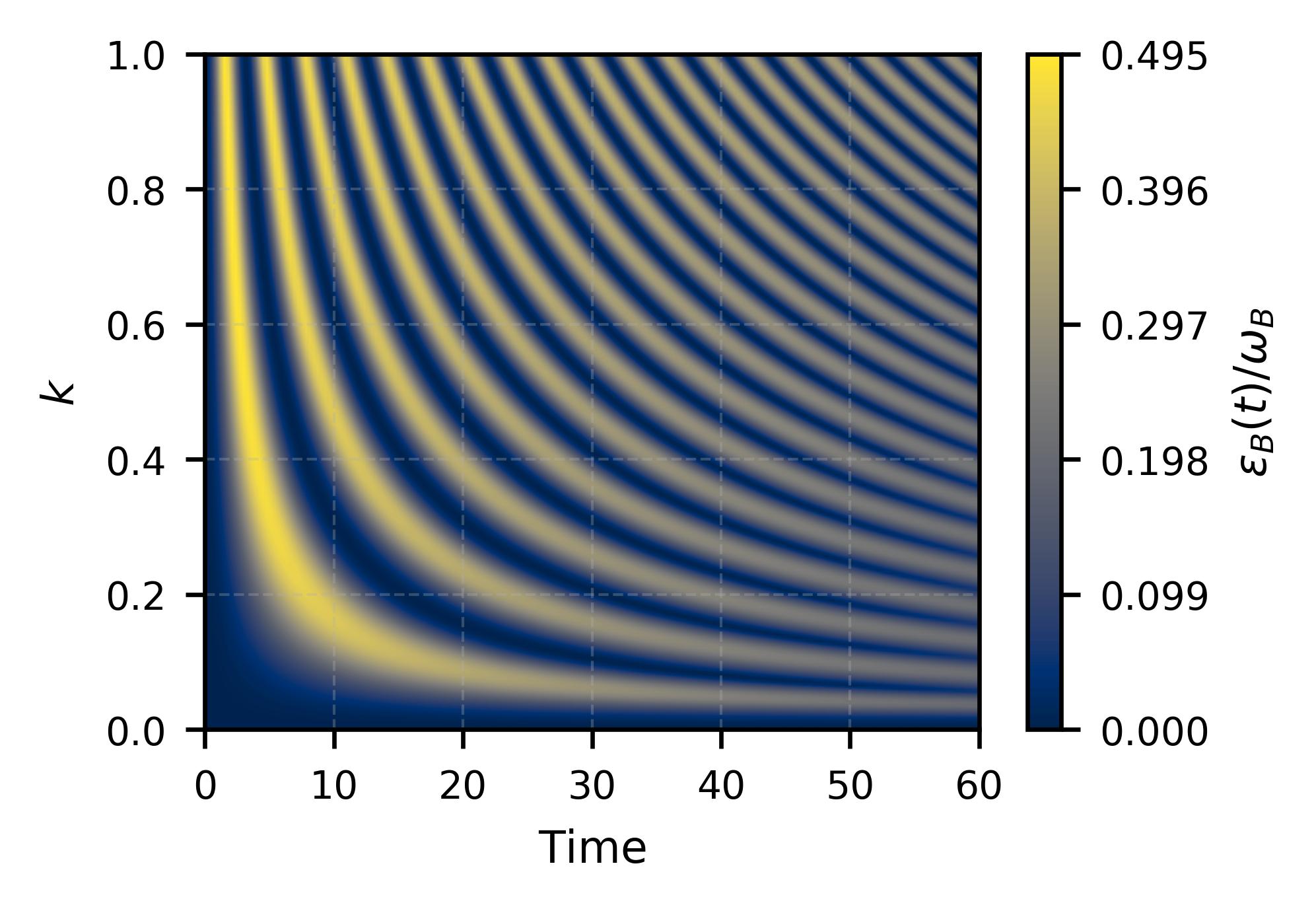}%
		}
		
		\caption{Density plots of trace distance derivative, relative entropy of coherence, normalized power of charging and normalized ergotropy, namely $\sigma_{B}(t)$, $C_B(t)$, $\Delta P_B (t)/\omega_{B}$, and $\epsilon_{B}(t)/\omega_{B}$ in Figs~\ref{subfig:TD}, \ref{subfig:COB}, \ref{subfig:Po} and \ref{subfig:ERGO}, respectively. The time and coupling strength $k$ between charger-battery are varied along the x-axis and y-axis, respectively. Moreover, note that $k$ is in unit of ($\omega_{M_2}/10$.)}	\label{NOM}	
	\end{figure*}
	%%%%%%%%%%%%%%%%%%%%%%%%%%%%%%%%%%%%%%%%%%%%%%%%%%%%%%%%%%%%%%%%%%%%%%%%%%%%%%%%%%%%%%%%%%%%%%%%%%%%%%%%%%%%%
	
	For $k = 0$ and $\sigma_{B}(t) = 0$, the battery is completely isolated from its environment and retains its information over time (Fig.(8a)). As the coupling $k$ increases, $\sigma_{B}(t)$ evolves from a monotonic regime to a non-monotonic one, oscillating between negative and positive values. This behavior indicates that the interaction between charger and battery induces a transition from Markovian dynamics (weak coupling) to non-Markovian dynamics (strong coupling), where information flows back into the quantum battery due to the presence of memory effects.\\
	
	The impact of non-Markovian dynamics on the coherence storage $C_B(t)$ in the quantum battery is remarkable (Fig.(8b)). For all values of $k$, i.e., $C_B(0) = 0$, the battery starts in an incoherent state $\hat{\rho}_B (0)=\ket{0}_B\bra{0}$. Over time, the coherence increases monotonically in the weak coupling regime and non-monotonically in the strong coupling regime. This transition in behavior is a consequence of the underlying dynamics: coherence evolves in a Markovian manner for weak coupling and in a non-Markovian fashion for strong coupling between charger and battery. Similarly, the power of charging for the quantum battery follows the same behavior as the variation of coherence (Fig.(8c)). It is maximal when $k$ is large and vice versa. Therefore,  the coupling between the charger and battery enhances the quantum battery power of charging. The maximum extracted work from the quantum battery is analyzed by the quantum ergotropy of the quantum battery (Fig.(8d)). The evolution of ergotropy is similar to the variation of its coherence. Moreover, the impact of non-Markovian dynamics on the ergotropy is remarkable due to the evolution of quantum battery coherence over time.\\

	As a result, the coupling between the quantum charger and the quantum battery is responsible for the backflow of information between the quantum charger and the quantum battery over time. It increases the degree of non-Markovianity, correlation, coherence, and power of charging. In addition, it has a very positive impact on the charging process of the quantum battery.

	%%%%%%%%%%%%%%%%%%%%%%%%%%%%%%%%%%%%%%%%%%%%%%%%%%%%%%%%%%%%%%%%%%%%%%%%%%%%%%%%%%%%%%%%%%%%%%%%%%%%%%%%%%%%%
	
	\section{Conclusion}\label{conc}

	In this work, we have explored the theoretical aspects that highlight the impact of a quantum autonomous thermal machine and coherence driving on the charging process. In the proposed model, we investigated the QATM composed of two qubits, \( M_{12} = M_1 \otimes M_2 \). Each of these qubits, i.e., \( M_1 \) and \( M_2 \), is coupled to  Markovian cold and hot reservoirs, \( R_1 \) and \( R_2 \). The role of coherence driving is represented by the injection of an external field into the quantum charger. The primary aim of this work is to connect the charger to the thermal machine, where this machine mediates the interaction between the charger, battery, and the reservoirs \( R_1 \) and \( R_2 \).\\
	
	The dynamic interaction between the quantum charger and QATM serves as a filter, mitigating the impact of decoherence effects from the reservoirs $R_1$ and $R_2$ on the composite system of the quantum charger-battery. The non-Markovianity in all subsystems is due to their exchange of correlations over time. Moreover, we showed the role of injecting an external field into the quantum charger and its impact on controlling its internal energy and preserving its coherence against the decoherence effects of the reservoirs. The interaction between the charger and QATM reduced the effect of reservoir-induced decoherence and played a positive role in the energy transfer between the external system components. The role of the external field (coherence driving) in the quantum charger is to preserve its internal energy, thereby enhancing the charging power of the quantum battery and increasing the maximal extractable work. Hence, we have obtained that the coherence driving raised the maximum ergotropy by approximately 40\% compared to the case without coherence driving.
	\\

	Finally, we examined the impact of the interaction between the quantum charger and quantum battery, where we fixed the coupling between QATM and the charger. This part showed us a remarkable dynamical effect on the quantum battery. Indeed, the robust coupling between the quantum charger and quantum battery is basically responsible for the transition from Markovian to non-Markovian dynamics. Hence, the robustness of this coupling increased the power of charging for the quantum battery, which gives rise to an enhancement of the charging capacity of the quantum battery, as well as an increase in the maximum extracted work from the quantum battery over time.\\
	
	Our work is closely related to several studies on non-Markovian effects in external systems and those of the generalized Landauer bound as discussed in our paper in \cite{intro10a}. In particular, our paper extended these effects to quantum coherence charging batteries. Moreover, the impact of Markovian channels on the capacity of quantum batteries in the case of Bell-diagonal states is discussed in \cite{intro11}. However, our work focused on controlling non-Markovian dynamics using various considerations of the coupling between the charger and quantum battery. This approach allowed us to study both Markovian and non-Markovian behaviors depending on whether the coupling between the charger and battery, or between the QATM and charger, is weak or strong, respectively.
	
	The role of coherence in quantum work extraction and energy storage under Markovian dynamics is examined in \cite{intro13}. We further investigate the role of coherence in the charging process using the relative entropy of coherence. Our results showed that it is possible to have a transfer of coherence between a quantum charger and battery under the exchange of correlations in non-Markovian dynamics. Moreover, our work demonstrated that the QATM is able to control the decoherence effects in the charger-battery system. Hence, this makes our contribution an extension of the previous works in the context of the QATM and coherence-charging batteries. Moreover, our work can be experimentally implemented within the framework of superconducting qubits since we have used experimentally feasible parameters, namely the energy spacing is taken in the GHz range, while the coupling strengths in the MHz range~\cite{sup1,sup2,sup3}.

	%%%%%%%%%%%%%%%%%%%%%%%%%%%%%%%%%%%%%%%%%%%%%%%%%%%%%%%%%%%%%%%%%%%%%%%%%%%%%%%%%%%%%%%%%%%
	\acknowledgments
	%%%%%%%%%%%%%%%%%
	A.~K acknowledges CNRST-Morocco support for this research within the Program " PhD-ASsociate Scholarship – PASS".

	%%%%%%%%%%%%%%%%%%%%%%%%%%%%%%%%%%%%%%%%%%%%%%%%%%%%%%%%%%%%%%%%%%%%%%%%%%%%%%%%%%%%%%%%%%%%
	\section*{Declaration of Interest}
	The authors declare that they have no conflict of interest.
	\section*{Data availability statement}
	No data statement is available.
	%----------------------------------------------------------------------------------------------------------


\begin{thebibliography}{9}
		%%%%%%%%%%%%%%%%%%%%%%%%%%%% INTRODUCTION %%%%%%%%%%%%%%%%%%%%%%%%%%%%
		\bibitem{intro1}
		Max F Riedel et al, "The European quantum technologies flagship programme
		" , Quantum Sci. Technol. \textbf{2} 030501 (2017).
		\bibitem{intro2}	
		Antonio Acín et al,"The quantum technologies roadmap: a European community view",  New J. Phys. \textbf{20} 080201 (2018).
		\bibitem{intro3}
		Giulia Gemme et al, "IBM Quantum Platforms: A Quantum Battery Perspective" Batteries 8(5), 43 (2022).
		\bibitem{intro4}
		Luca Razzoli et al, "Cyclic solid-state quantum battery: thermodynamic characterization and quantum hardware simulation"Quantum Sci. Technol. \textbf{10}  015064 (2025).
		\bibitem{intro5}
		J.Q. Quach et al, "Quantum batteries: The future of energy storage?". Joule, \textbf{7}(10), 2195-2200 (2023).
		\bibitem{intro6}
		G. Gour et al, "The resource theory of informational nonequilibrium in thermodynamics". Physics Reports \textbf{583}  1–58 (2015).
		\bibitem{intro7}
		F.Sapienza et al, "Correlations as a resource in quantum thermodynamics". Nat Commun \textbf{10}, 2492 (2019).
		\bibitem{intro8}
		John Goold et al, "The role of quantum information in thermodynamics—a topical review", J. Phys. A: Math. Theor. \textbf{49} 143001 (2016).
		\bibitem{intro9}
		Lipka-Bartosik et al. "Thermodynamic computing via autonomous quantum
		thermal machines", Sci. Adv. \textbf{10}, eadm8792 (2024).
		\bibitem{intro10}
		Ralph Silva et al, "Performance of autonomous quantum thermal machines: Hilbert space dimension as a thermodynamical resource", Phys. Rev. E \textbf{94}, 032120 (2016).
		\bibitem{intro10aa}
		A.~Canzio, V.~Cavina, R.~Menta, and V.~Giovannetti, 
			``Extracting and charging energy into almost unknown quantum states,'' 
			\textit{arXiv preprint} arXiv:2509.08899, 2025.
		\bibitem{intro10aaa}
		Zhi-Guang Lu et al, "Topological Quantum Batteries", Phys. Rev. Lett. 134, 180401 (2025).
		
		\bibitem{intro10a}
		A. Khoudiri et al, "Non-Markovianity and a generalized Landauer bound for a minimal quantum autonomous thermal machine with a work qubit", Phys. Rev. E \textbf{111}, 044124 (2025).
		
		\bibitem{into10b}
		A. Oularabi, A. El Allati, K. El Anouz, Enhancing  ergotropy of quantum batteries through coherence and non-Markovianity, Physica A: Statistical Mechanics and its Applications \textbf{679}, 131003 (2025).
		\bibitem{into10c}
		Ju-Yeon Gyhm, Uwe R. Fischer; 
		"eneficial and detrimental entanglement for quantum battery charging".
		AVS Quantum Sci. 6, 012001 (2024).		
		\bibitem{intro11}
		Wang, YK., Ge, LZ., Zhang, T. et al. Dynamics of quantum battery capacity under Markovian channels. Quantum Inf Process \textbf{24}, 34 (2025).
		\bibitem{intro11a}
			Z. Khodadad, M. Mahdian, and G. Hanna, ``Dark Excitonic State Preparation in a Symmetry‐Protected Open Quantum Battery,'' \textit{Advanced Quantum Technologies}, 2025, e00211.
		
		\bibitem{intro12} 
		Shastri, R., Jiang, C., Xu, GH. et al. Dephasing enabled fast charging of quantum batteries. npj Quantum Inf 11, 9 (2025). 
		\bibitem{intro13}
		Salvatore Tirone et al, "Quantum work extraction efficiency for noisy quantum batteries: The role of coherence", Phys. Rev. A \textbf{111}, 012204 (2025).
		\bibitem{intro14}
		Jonatan Bohr Brask et al, "Autonomous quantum thermal machine for generating steady-state entanglement",  New J. Phys. 17 113029 (2015).
		\bibitem{intro15}
		Gonzalo Manzano et al, "Autonomous thermal machine for amplification and control of energetic coherence", Phys. Rev. E \textbf{99}, 042135 (2019).
		\bibitem{intro16}
		Noah Linden et al, "How Small Can Thermal Machines Be? The Smallest Possible Refrigerator", Phys. Rev. Lett. \textbf{105}, 130401 (2010).
		\bibitem{intro17}
		Nicolas Brunner et al , "Virtual qubits, virtual temperatures, and the foundations of thermodynamics", Phys. Rev. E \textbf{85}, 051117 (2012).
		\bibitem{intro18}
		Paolo Abiuso et al, "Non-Markov enhancement of maximum power for quantum thermal machines", Phys. Rev. A \textbf{99}, 052106 (2019).
		\bibitem{intro19}
		El Allati et al, " Non-Markovian effects on the performance of a quantum Otto refrigerator". Physics Letters A, \textbf{496}, 129316 (2024).
		\bibitem{intro19a}
		Caravelli, F., Yan, B., García-Pintos, L. P., \& Hamma, A. (2021).
			Energy storage and coherence in closed and open quantum batteries.
			\textit{Quantum}, \textbf{5}, 505.
		\bibitem{intro20}
		Tsakmakidis et al. "Quantum coherence–driven self-organized criticality and nonequilibrium light localization", Sci. Adv.\textbf{ 4}: eaaq0465 (2018).
		\bibitem{intro21}
		R. R. Rodríguez et al , "Catalysis in charging quantum batteries " , Phys. Rev. A \textbf{107}, 042419 (2023).
		\bibitem{intro22}
		Shastri, R., Jiang, C., Xu, GH. et al. "Dephasing enabled fast charging of quantum batteries". npj Quantum Inf \textbf{11}, 9 (2025). 
		\bibitem{intro23}
		Alba Crescente et al, "Analytically Solvable Model for Qubit-Mediated Energy Transfer between Quantum Batteries". Entropy, \textbf{25}, 758 (2023).
		\bibitem{intro24}
		Gian Marcello Andolina et al,"Charger-mediated energy transfer in exactly solvable models for quantum batteries
		", Phys. Rev.B \textbf{98}, 205423 (2018).
		
		\bibitem{intro25}
		Gian Marcello Andolina et al, "Extractable Work, the Role of Correlations, and Asymptotic Freedom in Quantum Batteries", Phys. Rev. Lett. \textbf{122}, 04770 (2019).
		\bibitem{intro26}
		Donato Farina et al, "Charger-mediated energy transfer for quantum batteries: An open-system approach", Phys. Rev. \textbf{B} 99, 035421 (2019).
		\bibitem{intro26a}
			Sathe, P., \& Caravelli, F. (2025).
			Universally charging protocols for quantum batteries: A no-go theorem.
			\textit{Europhysics Letters}, \textbf{150}(4), 48001.
		%%%%%%%%%%%%%%%%%%%%%%%%%%%%QATM-Charger-Battery Model%%%%%%%%%%%%%%%%%%%%%%%%%%%%	
		\bibitem{MODEL1}
		Nicolas Brunner et al. Phys. Rev. E \textbf{85}, 051117 (2012).
		\bibitem{MODEL2} 	
		S.Lorenzo et al, "Composite quantum collision models", Phys. Rev. A, \textbf{96}, 032107 (2017).
		\bibitem{MODEL3}
		H.-P.Breuer, F.Petruccione. "The theory of open quantum systems". Oxford University Press(2002).
		\bibitem{MODEL4}
		A.Rivas Vargas, "Open quantum systems and quantum information dynamics", (Doctoral thesis, Universitaire Ulm) (2011).
		
		\bibitem{ModelAch1}
		A.~Khoudiri, K.~El~Anouz, and A.~El~Allati,
		``Correlation enhanced autonomous quantum battery charging via structured reservoirs,''
		arXiv:2508.18086 [quant-ph] (2025).
		\bibitem{ModelAch2}
		A.~Khoudiri, K.~El~Anouz, and A.~El~Allati,
		``Generation of quantum entanglement in autonomous thermal machines: Effects of non-Markovianity, Hilbert space structure, and quantum coherence,''
		arXiv:2508.18056 [quant-ph] (2025).
		%%%%%%%%%%%%%%%%%%%%%%%%%%%%NON_Markovianity%%%%%%%%%%%%%%%%%%%%%%%%%%%%	
		\bibitem{NONMARKOVIANITY1}
		Steve Campbell at al, "Precursors of non-Markovianity". New J. Phys. \textbf{21}, 053036 (2019). 
		\bibitem{NONMARKOVIANITY2}
		Bassano Vacchini, "Non-Markovian master equations from piecewise dynamics".	Phys. Rev. A \textbf{87}, 030101(R) (2013).
		\bibitem{NONMARKOVIANITY3}
		S. Haseli et al, "Non-Markovianity through flow of information between a system and an environment". Phys. Rev. A \textbf{90}, 052118 (2014).
		\bibitem{NONMARKOVIANITY4}
		F.F.Fanchini et al, "Non-Markovianity through Accessible Information". Phys. Rev. Lett. \textbf{112}, 210402 (2014).
		\bibitem{NONMARKOVIANITY5}
		Asutosh Kumar, "Multiparty quantum mutual information: An alternative definition
		". Phys. Rev. A\textbf{96}, 012332 (2017).
		\bibitem{sup1}
		H.~K.~Warner, J.~Holzgrafe, B.~Yankelevich, D.~Barton, S.~Poletto, C.~J.~Xin, \textit{et al.},
		``Coherent control of a superconducting qubit using light,''
		\textit{Nat. Phys.}, 1--8 (2025).
		\bibitem{sup2}
		A.~Anferov, F.~Wan, S.~P.~Harvey, J.~Simon, and D.~I.~Schuster,
		``Millimeter-wave superconducting qubit,''
		\textit{PRX Quantum} \textbf{6}, 020336 (2025).
		\bibitem{sup3}
		P.~M.~Harrington, M.~Li, M.~Hays, W.~Van~De~Pontseele, D.~Mayer, H.~D.~Pinckney, \textit{et al.},
		``Synchronous detection of cosmic rays and correlated errors in superconducting qubit arrays,''
		\textit{Nat. Commun.} \textbf{16}, 6428 (2025).	
	\end{thebibliography}
\end{document}